%% file: ms.tex
\documentclass[12pt,a4paper]{article}
\usepackage{a4wide}
\usepackage{moresize}
\usepackage{amsmath}
\usepackage{amssymb}
\usepackage{amsfonts}
\usepackage{epsfig}
\usepackage{exscale}
\usepackage{float}
\graphicspath{{figures/}}
\usepackage{bbm}
\usepackage[numbers,sort&compress]{natbib}

\newcommand{\Z}{{\mathbb{Z}}}

\newcommand{\CP}{{\mathbb{C}P}}

\makeatletter
\@addtoreset{equation}{section}
\makeatother

\usepackage{booktabs} 
\usepackage{tikz} 
\usetikzlibrary{shapes.multipart}
\usetikzlibrary{calc}
\DeclareMathOperator{\Tr}{Tr}
\newcommand{\ket}[1]{\left| #1 \right>} 
\newcommand{\matrixel}[3]{\left< #1 \vphantom{#2#3} \right| #2 \left| #3 \vphantom{#1#2} \right>} 

\title{$SU(3)$ Quantum Spin Ladders as a Regularization of the $\CP(2)$ Model 
at Non-Zero Density: \\ From Classical to Quantum Simulation}

\author{W.\ Evans, U.\ Gerber, M.\ Hornung, and U.-J.\ Wiese
\footnote{Contact information: M.\ Hornung, hornung@itp.unibe.ch, 
+41 31 613 5460.}
\\ \\
\small Albert Einstein Center for Fundamental Physics \\
\small Institute for Theoretical Physics, University of Bern \\
\small Sidlerstrasse 5, CH-3012 Bern, Switzerland}

\begin{document} 

\maketitle

\vspace{-1cm}

\begin{abstract} \normalsize

Quantum simulations would be highly desirable in order to investigate the finite
density physics of QCD. $(1+1)$-d $\CP(N-1)$ quantum field theories are toy 
models that share many important features of QCD: they are asymptotically free, 
have a non-perturbatively generated massgap, as well as $\theta$-vacua. $SU(N)$ 
quantum spin ladders provide an unconventional regularization of 
$\CP(N-1)$ models that is well-suited for quantum simulation with ultracold 
alkaline-earth atoms in an optical lattice. In order to validate
future quantum simulation experiments of $\CP(2)$ models at finite density, here
we use quantum Monte Carlo simulations on classical computers to investigate
$SU(3)$ quantum spin ladders at non-zero chemical potential. This reveals a
rich phase structure, with single- or double-species Bose-Einstein ``condensates'', 
with or without ferromagnetic order.

\end{abstract}

\newpage
 
\section{Introduction}

Monte Carlo simulations of Wilson's lattice QCD \cite{Wil74} are very 
successful in addressing static properties of hadrons \cite{Dur08,Baz10}
as well as the equilibrium thermodynamics of quarks and gluons at zero baryon 
density \cite{Aok06,Baz09}. The real-time dynamics and the non-zero density 
physics of QCD \cite{Raj01}, on the other hand, remain largely unexplored, because Monte 
Carlo simulations then suffer from very severe sign and complex action problems.
Quantum simulation experiments are very promising for addressing these
challenging questions, because quantum hardware (whose dynamics naturally 
incorporates quantum entanglement) does not suffer from such
problems \cite{Fey82,Cir12,Llo96,Jak98, Lew12, Blo12}.
Indeed, quantum simulation experiments have already been carried out successfully in the
context of condensed matter physics. In particular, the real-time evolution
through a quantum phase transition in the bosonic Hubbard model, which 
separates a Mott insulator from a superfluid, has been realized in quantum 
simulation experiments with ultracold bosonic atoms in an optical lattice 
\cite{Gre02}. Similar experiments with fermionic atoms aim at quantum 
simulations of the fermionic Hubbard model, in the context of high-temperature
superconductivity. The current experiments with fermionic gases have not 
yet succeeded to reach sufficiently low temperatures to explore the possible 
existence of high-temperature superconductivity in the fermionic Hubbard model.
However, medium-range antiferromagnetic correlations have already been observed
\cite{Maz17}.

These impressive developments in the quantum simulation of condensed matter
systems provide a strong motivation to explore the feasibility of quantum
simulation experiments of QCD and other quantum field theories relevant in
particle physics. While it seems difficult to embody Wilson's lattice 
QCD in ultracold quantum matter, an attractive alternative lattice 
regularization of QCD and other asymptotically free field theories is provided
by quantum link models \cite{Hor81, Orl90, Cha97, Bro99, Bro04}.
Quantum links are generalized quantum spins
(associated with the links of a lattice) with an exact gauge symmetry.
Wilson's link variables are classical $SU(3)$-valued parallel transporter 
matrices with an infinite-dimensional link Hilbert space. $SU(3)$ quantum links
are again $3 \times 3$ matrices, but their matrix elements are non-commuting
operators that act in a finite-dimensional link Hilbert space. This makes 
quantum link models ideally suited for quantum simulation experiments in which a
finite number of quantum states of ultracold matter can be controlled 
successfully \cite{Wie13}. Indeed, quantum simulation experiments of Abelian 
\cite{Tag12,Zoh12,Ban12,Kas17} and
non-Abelian gauge theories \cite{Tag13,Zoh13,Ban13}, some based on  
quantum link models have already been proposed. In 
particular, ultracold alkaline-earth atoms in an optical superlattice \cite{Ban13} 
are natural physical objects that can embody non-Abelian $U(N)$ and $SU(N)$ 
gauge theories. 

While first quantum simulation experiments of relatively simple Abelian and 
non-Abelian lattice gauge theories are expected in the near future, the quantum 
simulation of QCD remains a long-term goal \cite{Wie14}. The quantum link 
regularization of QCD \cite{Bro99} involves an additional spatial dimension (of short
physical extent) in which the discrete quantum link variables form emergent 
continuous gluon fields via dimensional reduction. The extra dimension also 
gives rise to naturally light domain wall quarks with an emergent chiral 
symmetry. Incorporating these important dynamical features in quantum 
simulation experiments will be challenging, but does not seem impossible. In 
particular, synthetic extra dimensions have already been realized in quantum 
simulation experiments with alkaline-earth atoms \cite{Man15}.

In order to explore the feasibility of quantum simulation experiments of 
QCD-like theories, it is natural to investigate $(1+1)$-d $\CP(N-1)$ models 
\cite{Add78, Eic78}. These quantum field theories share crucial features with QCD: they are 
asymptotically free, have a non-perturbatively generated massgap, as well as 
non-trivial topology and hence $\theta$-vacuum states. In particular, the 
$\CP(N-1)$ model has a global $SU(N)$ symmetry that gives rise to interesting 
physics at non-zero density, which can be explored via chemical potentials. As 
in QCD, the direct classical simulation of $\CP(N-1)$ model $\theta$-vacua, 
finite density physics, or dynamics in real-time suffer from severe sign 
problems, and thus strongly motivate the need for quantum simulation.
\footnote{It should be noted that classical sign-problem-free simulations of 
$\CP(N-1)$ models at non-zero density are possible after an analytic rewriting 
of the partition function \cite{Wol10,Bru15,Rin17}. Unfortunately, this does not seem to extend 
to QCD.} Again, alkaline-earth atoms in an optical superlattice are natural 
degrees of freedom to realize the $SU(N)$ symmetry of $\CP(N-1)$ models \cite{Laf16, Laf15}.

In complete analogy to the quantum link regularization of QCD, $(1+1)$-d 
$\CP(N-1)$ models can be regularized using $(2+1)$-d $SU(N)$ quantum spin 
ladders \cite{Bea05, Bea06}. Again, there is an extra spatial dimension of short physical extent in 
which the discrete quantum spins form emergent continuous $\CP(N-1)$ fields via 
dimensional reduction. The continuum limit of the $(1+1)$-d $\CP(N-1)$ quantum
field theory is taken by gradually increasing the extent $L'$ of the extra
dimension. Thanks to asymptotic freedom, this leads to an exponential increase 
of the correlation length $\xi \gg L'$ in the physical dimension, and thus to
dimensional reduction from $(2+1)$-d to $(1+1)$-d, similar to the $O(3)=\CP(1)$ 
model \cite{Cha88,Has91}. All this is 
analogous to QCD, but has the great advantage that it can already be 
investigated with currently available experimental quantum simulation 
techniques. In particular, by varying $L'$ it should be possible to approach
the continuum limit of the $(1+1)$-d $\CP(N-1)$ quantum field theory in 
ultracold atom experiments.

In order to validate and support upcoming quantum simulation experiments of 
$\CP(N-1)$ models at zero and non-zero density, here we use quantum Monte Carlo 
calculations (on classical computers) to simulate $SU(N)$ quantum spin ladders 
with chemical potential. This method was first used in \cite{Cha02} to investigate
the $(1+1)$-d $O(3)=\CP(1)$ model at non-zero chemical potential using a meron-cluster
algorithm. Here we focus on the $N = 3$ case of the $\CP(2)$ model with
a global $SU(3)$ symmetry \cite{Eva16} which is accessible to quantum simulation 
experiments. Since $SU(3)$ has rank 2, with two commuting generators $T^3$ and 
$T^8$, there are two independent chemical potentials $\mu_3$ and $\mu_8$. A 
chemical potential $\mu_3 \neq 0$ generically breaks the global $SU(3)$ symmetry
explicitly down to $U(1)_3 \times U(1)_8$. As we will show, at zero temperature 
the $U(1)_3$ symmetry undergoes the Kosterlitz-Thouless phenomenon and the 
remaining symmetry is reduced to $U(1)_8$. Due to the Mermin-Wagner theorem, 
this is as close as a $(1+1)$-d quantum field theory can come to Bose-Einstein 
``condensation''. Interestingly, for $\mu_3 = 0$ and $\mu_8 \neq 0$, the 
$SU(3)$ symmetry is explicitly broken only to $SU(2)_{123} \times U(1)_8$. Now 
the $U(1)_8$ symmetry undergoes the Kosterlitz-Thouless phenomenon. The 
$SU(2)_{123}$ symmetry then gives rise to a double-species Bose-Einstein 
``condensate''. The $SU(2)_{123}$ ``spin'' $(T^1,T^2,T^3)$ is a conserved order 
parameter, just like the total spin of a ferromagnetic quantum spin chain. 
Remarkably, we thus obtain a double-species ``ferromagnetic'' Bose-Einstein 
``condensate''. It will be most exciting to explore this rich phase structure of 
the $(1+1)$-d $\CP(2)$ model with experimental quantum simulations of 
alkaline-earth atoms in an optical lattice. Our classical simulations can serve 
as a valuable tool to validate future experiments of this kind.

The rest of the paper is organized as follows. In Section 2 we introduce 
$(2+1)$-d $SU(N)$ quantum spin ladders as a regularization of $(1+1)$-d 
$\CP(N-1)$ models. In particular, we discuss the mechanism of dimensional 
reduction from $(2+1)$-d to $(1+1)$-d. In Section 3 we focus on $(2+1)$-d 
$SU(3)$ quantum spin ladders and the resulting $(1+1)$-d $\CP(2)$ model, with a 
special emphasis on its finite-density physics. We present results of quantum 
Monte Carlo calculations to explore the phase diagram and to investigate the 
nature of the various phases. We also measure various correlation 
functions in order to investigate the properties of the excitations in 
different regions of the phase diagram. Finally, Section 4 contains our 
conclusions. The details of a quantum Monte Carlo worm algorithm are 
described in an appendix.

\section{From $SU(N)$ Quantum Spin Ladders to \\
$\CP(N-1)$ Models}
\label{sec:su3system}

In this section, we introduce an antiferromagnetic $SU(N)$ quantum spin ladder
and discuss its low-energy effective field theory, which leads to the
$\CP(N-1)$ model via dimensional reduction.
 
\subsection{Antiferromagnetic $SU(N)$ Quantum Spin Ladder}

\begin{figure}[tbh]
\begin{center}
\includegraphics[width=\columnwidth]{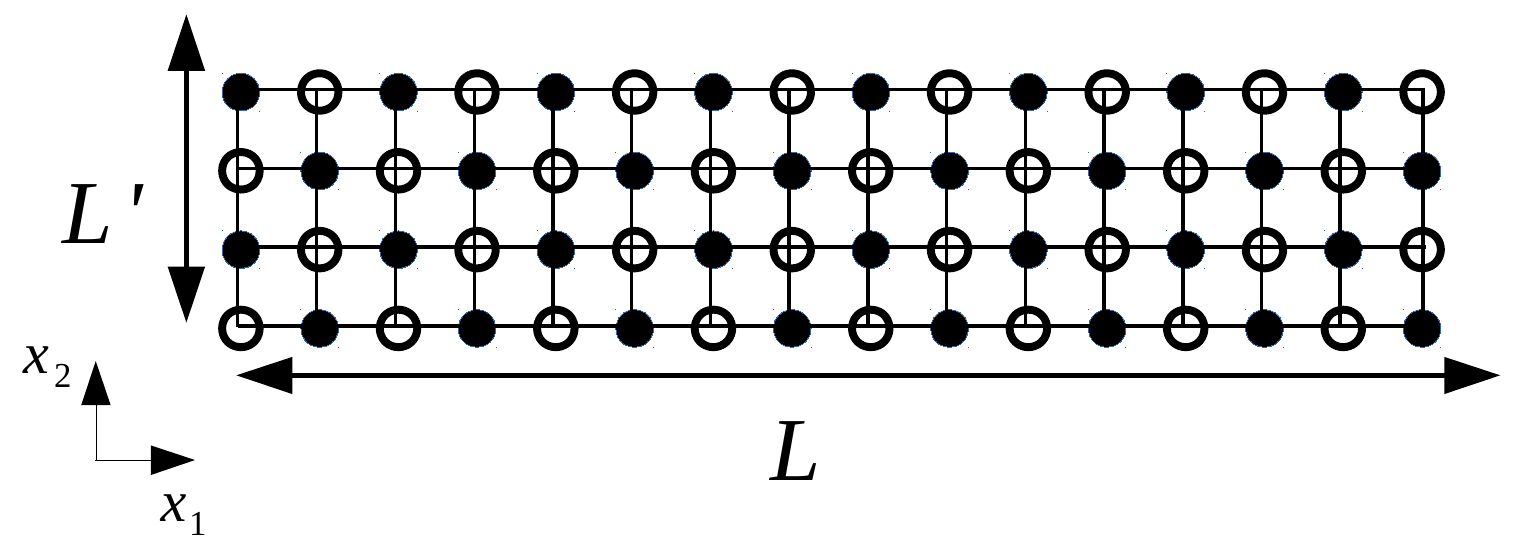}
\caption{\it Spin ladder of length $L$ in the 1-direction, 
consisting of $n = L'/a$ transversely coupled spin chains. The even 
sites on sublattice $A$ (open circles) carry an $SU(N)$ quantum spin
in the fundamental $\{N\}$ representation, while the odd sites on
sublattice $B$ (filled circles) carry the anti-fundamental 
$\{\overline{N}\}$ representation. Periodic and open boundary
conditions are imposed in the 1- and 2-direction, respectively.}
\label{ladder}
\end{center}
\end{figure}

\begin{figure}[tbh]
\begin{center}
\includegraphics[
width=0.7\columnwidth]{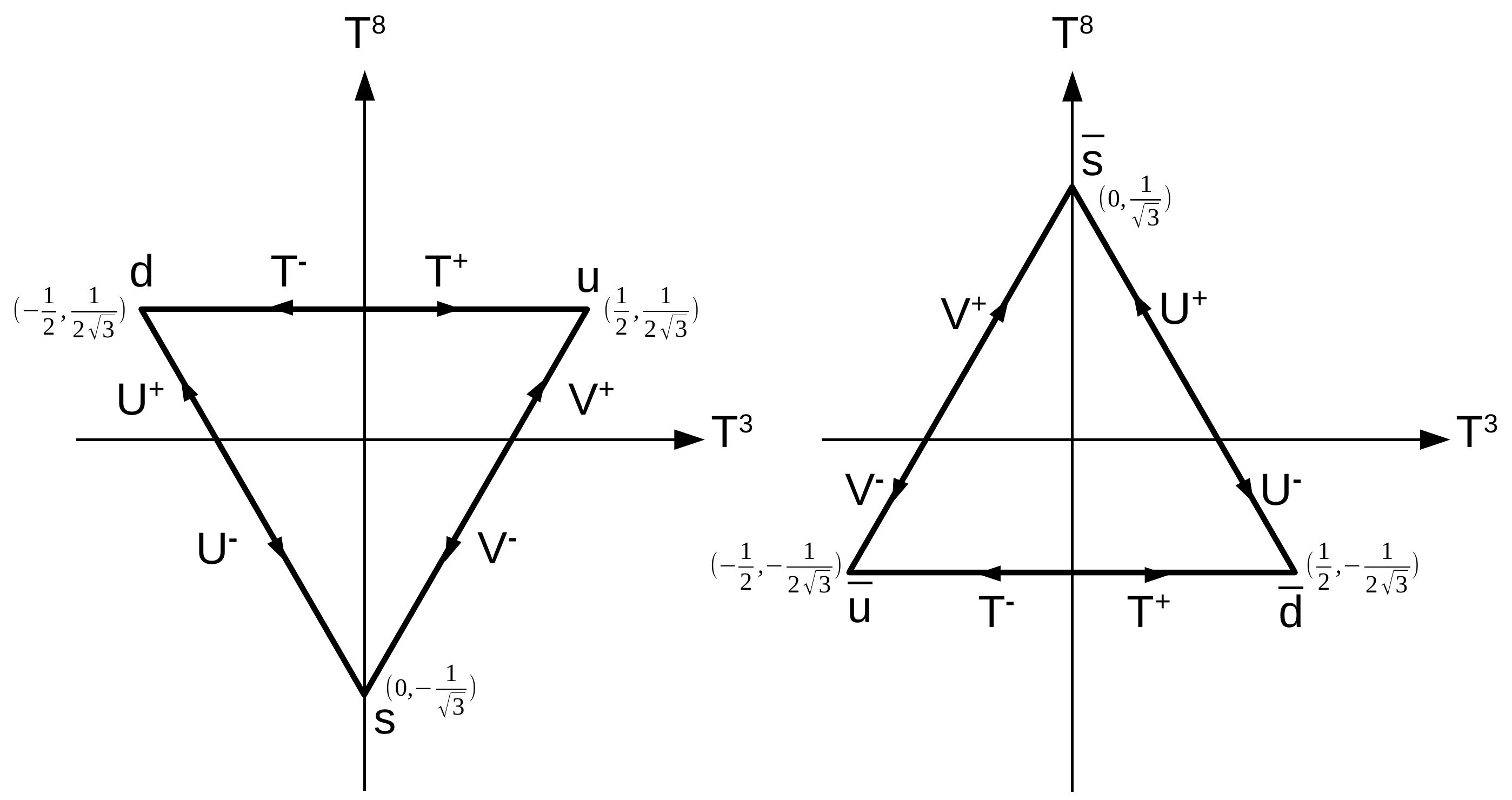}
\caption{\it Fundamental $SU(3)$ triplet representation $\{3\}$ (left) and 
anti-fundamental representation $\{\overline{3}\}$ (right). The shift 
operators $T^\pm$, $U^\pm$, and $V^\pm$ (cf.\ eq.(\ref{shiftoperators}))
relate the different states of the multiplets.}
\label{33bar}
\end{center}
\end{figure}

Let us consider a 2-d bipartite square lattice of large extent $L$ in the 
periodic 1-direction and short extent $L'$ in the 2-direction with open boundary
conditions. In the continuum limit, the 2-direction will ultimately disappear
via dimensional reduction, while the 1-direction remains as the physical spatial
dimension \cite{Bea05}. As illustrated in Fig.\ref{ladder}, we distinguish the sites 
$x \in A$ of the even sublattice $A$ from the neighboring sites $y \in B$ of the
odd sublattice $B$. We place $SU(N)$ quantum spins $T^a_x = \frac{\lambda^a}{2}$
in the fundamental representation $\{N\}$ on sublattice $A$. Here $\lambda^a$ 
(with $a \in \{1,2,\dots,N^2-1\}$) are the generators of the $SU(N)$ algebra;
for $N = 3$ they are the Gell-Mann matrices. As a consequence, the spins obey 
the commutation relations
\begin{equation}
[T_x^a,T_{x'}^b] = i \delta_{xx'} f_{abc} T_x^c \ ,
\end{equation}
with the structure constants $f_{abc}$ of the $SU(N)$ algebra. For $N = 3$ it is
useful to introduce the shift operators
\begin{equation}
\label{shiftoperators}
T^\pm = T^1 \pm i T^2, \quad V^\pm = T^4 \pm i T^5, \quad 
U^\pm = T^6 \pm i T^7 \ .
\end{equation}
On sublattice $B$, on the other hand, we place quantum spins 
$\overline{T}^a_y = - \frac{{\lambda^a}^*}{2}$ in the complex conjugate 
anti-fundamental representation $\{\overline{N}\}$, which again obey
\begin{equation}
[\overline{T}_y^a,\overline{T}_{y'}^b] = i \delta_{yy'} f_{abc} \overline{T}_y^c 
\ ,
\end{equation}
For $SU(3)$ the weight diagrams of the fundamental and anti-fundamental representations $\{3\}$ and
$\{\overline{3}\}$ are illustrated in Fig.\ref{33bar}.

An antiferromagnetic $SU(N)$ quantum spin ladder (with $J > 0$) is then 
described by the nearest-neighbor Hamiltonian
\begin{equation}
\label{Hamiltonian}
H = J \sum_{\langle xy \rangle} T^a_x \overline{T}^a_y \ ,
\end{equation}
which commutes with the total $SU(N)$ spin
\begin{equation}
T^a = \sum_{x \in A} T^a_x + \sum_{y \in B} \overline{T}^a_y \ , 
\end{equation}
i.e.\ $[H,T^a] = 0$.

Let us also couple a chemical potential $\mu_a$ to the conserved non-Abelian
$SU(N)$ charge $T^a$ by subtracting a term $\mu_a T^a$ from the Hamiltonian. 
For $N = 3$, such a term can always be diagonalized to $\mu_3 T^3 + \mu_8 T^8$
by a unitary transformation, with two independent chemical potentials $\mu_3$ 
and $\mu_8$. The grand canonical partition function (at inverse temperature 
$\beta$) then takes the form
\begin{equation}
Z = \mbox{Tr} \exp(- \beta (H - \mu_a T^a )) \ .
\end{equation}

Let us first consider the $SU(N)$ spin system at zero temperature,
$\beta \rightarrow \infty$, in the infinite-volume limit, 
$L, L' \rightarrow \infty$. For $N \leq 4$ the system then breaks its $SU(N)$
symmetry down to $U(N-1)$ \cite{Har03}. This generalizes the 
$SU(2) \rightarrow U(1)$ symmetry breaking of the antiferromagnetic Heisenberg 
model to $SU(3)$ and $SU(4)$ quantum spin systems. For $N \geq 5$, on the other
hand, the simple Hamiltonian of eq.(\ref{Hamiltonian}) gives rise to a 
dimerized state with spontaneously broken lattice translation symmetry 
\cite{Har03}. One can easily imagine that more complicated Hamiltonians would
still support spontaneous $SU(N) \rightarrow U(N-1)$ breaking even for 
$N \geq 5$, which is what we will assume when we discuss general $N$. Later, we 
will focus our attention on the $N = 3$ case, in which the simple Hamiltonian of
eq.(\ref{Hamiltonian}) gives rise to spontaneous $SU(3) \rightarrow U(2)$
symmetry breaking.

\subsection{The $(2+1)$-d $\CP(N-1)$ Model as a Low-Energy
Effective Field Theory}

When a global $G = SU(N)$ symmetry breaks spontaneously to an $H = U(N-1)$ 
subgroup, $N^2 - 1 - (N-1)^2 = 2N$ massless Goldstone bosons arise. Their 
low-energy dynamics are described by an effective field theory in terms of
Goldstone boson fields $P(x)$ which take values in the coset space 
$G/H = SU(N)/U(N-1) = \CP(N-1)$. Here $x$ is a point in the $(2+1)$-d Euclidean
space-time continuum and $P(x)$ is an $N \times N$ matrix-valued field that
obeys
\begin{equation}
\label{relations}
P(x)^\dagger = P(x), \quad P(x)^2 = P(x), \quad \mbox{Tr} P(x) = 1 \ ,
\end{equation}
i.e.\ $P(x)$ is a Hermitean projection operator. The field $P(x)$ can be 
diagonalized by a unitary transformation $U(x) \in SU(N)$ such that
\begin{equation}
U(x) P(x) U(x)^\dagger = \mbox{diag}(1,0,\dots,0) \ .
\end{equation}
Since $P(x)$ is a projection operator with trace 1, the resulting diagonal 
matrix has one entry 1 and $N - 1$ entries 0. This matrix commutes with all
$U(N-1) = SU(N-1) \times U(1)$ matrices. Consequently, $P(x)$ is affected only
by those $SU(N)$ matrices $U(x)$ that belong to the coset space
$SU(N)/U(N-1) = \CP(N-1)$. Global symmetry transformations $\Omega \in SU(N)$,
which manifest themselves as
\begin{equation}
P(x)' = \Omega P(x) \Omega^\dagger \ ,
\end{equation}
indeed leave the defining relations of eq.(\ref{relations}) invariant.

The action of the low-energy effective theory contains all terms that respect
all symmetries of the underlying microscopic Hamiltonian of 
eq.(\ref{Hamiltonian}), in particular, the global $SU(N)$ symmetry. The leading
term in a systematic low-energy expansion has only two derivatives and gives
rise to the effective action
\begin{equation}
S[P] = \int d^3x \frac{\rho_s}{c} 
\mbox{Tr}\left[\partial_\mu P \partial_\mu P \right] \ .
\end{equation}
Here $\rho_s$ is the spin stiffness, $c$ is the spinwave velocity, and
$x_3 = c t$ is the appropriately rescaled Euclidean time coordinate. Note 
that we put $\hbar$ (but not $c$) to 1.

\subsection{Dimensional Reduction to the $(1+1)$-d 
$\CP(N-1)$ Model}

As long as we stay in the infinite-volume limit of the $(2+1)$-d Euclidean
space-time volume, the action from above describes strictly massless Goldstone
bosons. As soon as we deviate from this limit, for example, by making the extent
$L'$ of the 2-direction finite, the Mermin-Wagner theorem implies that the
continuous global $SU(N)$ symmetry can no longer break spontaneously. As a
consequence, the Goldstone bosons pick up an exponentially small mass \cite{Cha88,Has91, Bea05}. 
Interestingly, this non-perturbative effect is still captured by the Goldstone
boson effective action. Let us now consider a finite space-time volume, with 
periodic boundary conditions in the 1- and 3-directions and with open boundary
conditions in the 2-direction. For the moment we assume even extents $L$ and 
$L'$ of the 1- and 2-directions in the underlying $SU(N)$ quantum spin system. 
The low-energy effective action then takes the form
\begin{eqnarray}
S[P]&=&\int_0^{\beta c} dx_3 \int_0^L dx_1 \int_0^{L'} dx_2 \frac{\rho_s}{c} 
\mbox{Tr}\left[\partial_\mu P \partial_\mu P \right] \nonumber \\
&=&\int_0^\beta dt \int_0^L dx_1 \int_0^{L'} dx_2 \rho_s \mbox{Tr}\left[
\partial_i P \partial_i P + \frac{1}{c^2} \partial_t P \partial_t P\right] \ .
\end{eqnarray}
Let us now assume that both $L$ and $\beta$ are very large, while $L' = n a$ is
much smaller. Here $a$ is the lattice spacing of the underlying quantum spin 
system and $n$ is an even integer. As a consequence of the Mermin-Wagner 
theorem, the Goldstone bosons then pick up a non-zero mass $m$, which manifests 
itself as a finite spatial correlation length $\xi = 1/(m c)$. Let us first
assume that $\xi \gg L'$, which we will confirm later. Then the physics becomes
effectively independent of the short 2-direction and the system undergoes 
dimensional reduction from $(2+1)$-d to $(1+1)$-d. The $(1+1)$-d effective
action results from integrating over the 2-direction and by dropping terms that 
contain derivatives $\partial_2$, and takes the form
\begin{eqnarray}
S[P]&=&\int_0^\beta dt \int_0^L dx_1 L' \rho_s \mbox{Tr}\left[
\partial_1 P \partial_1 P + \frac{1}{c^2} \partial_t P \partial_t P\right]
\nonumber \\
&=&\int_0^{\beta c} dx_3 \int_0^L dx_1 \frac{1}{g^2} \mbox{Tr}\left[
\partial_1 P \partial_1 P + \partial_3 P \partial_3 P\right] \ .
\end{eqnarray}
This is the action of a $(1+1)$-d $\CP(N-1)$ quantum field theory with the
dimensionless coupling constant
\begin{equation}
\frac{1}{g^2} = \frac{L' \rho_s}{c} \ .
\end{equation}
Thanks to the asymptotic freedom of the $(1+1)$-d $\CP(N-1)$ model, the
correlation length is exponentially large in $1/g^2$, i.e.
\begin{equation}
\label{xi}
\frac{\xi}{L'} = 
C_N (\beta_1 g^2)^{\beta_2/(2 \beta_1^2)} \exp\left(\frac{1}{2 \beta_1 g^2}\right) = 
C_N \left(\frac{c N}{8 \pi L' \rho_s} \right)^{2/N} 
\exp\left(\frac{4 \pi L' \rho_s}{c N}\right) \ .
\end{equation}
Note that the correlation length $\xi$ is expressed in units of the extent $L'$ 
of the extra dimension (which has ultimately disappeared via dimensional 
reduction). The parameters $\beta_1 = N/(8 \pi)$ and $\beta_2 = N/(4 \pi)^2$ are
the 1- and 2-loop coefficients of the corresponding $\beta$-function. The value
of the $N$-dependent dimensionless constant $C_N$ determines the 
non-perturbatively generated massgap of the $(1+1)$-d $\CP(N-1)$ model in units 
of the scale $\Lambda_{\overline{MS}}$. This scale results via dimensional 
transmutation in the minimal modified subtraction scheme of dimensional 
regularization. Thanks to the knowledge of the exact S-matrix, which results 
from an infinite hierarchy of symmetries, the constants $C_N$ are analytically
known in $(1+1)$-d $O(N)$ models and other $(1+1)$-d asymptotically free field 
theories \cite{Bro04}. For $\CP(N-1)$ models, on the other hand, an exact 
S-matrix is not available because the hierarchy of symmetries exists only at the
classical level and is explicitly anomalously broken by quantum effects. As a 
result, determining $C_N$ requires numerical simulations. Eq.(\ref{xi}) indeed 
justifies the assumption that $\xi \gg L'$ already for moderately large values 
of $L'$. Dimensional reduction hence results as a consequence of asymptotic 
freedom. This is, in fact, completely analogous to how the continuum limit is
approached in the quantum link regularization of QCD \cite{Bro99}.

Let us now consider an odd extent $n = L'/a$ of the 2-direction for the 
underlying $SU(N)$ quantum spin ladder. Interestingly, there is a qualitative
difference between even and odd $n$. While even $n$ leads to the $\CP(N-1)$ 
model at vacuum angle $\theta = 0$, as we will discuss now, odd $n$ implies 
$\theta = \pi$ \cite{Bea05}. This is analogous to Haldane's conjecture for 
antiferromagnetic $SU(2)$ spin chains \cite{Hal83}. For integer spin 
these have a gap and are associated with the $(1+1)$-d $\CP(1)$ model at 
$\theta = 0$, while for half-integer spin they are gapless and correspond to 
$\theta = \pi$. For odd $n$, in the action that results after dimensional 
reduction, there is an additional topological term $i \theta Q[P]$, where
\begin{eqnarray}
Q[P]&=&\frac{1}{2 \pi i} \int_0^\beta dt \int_0^L dx_1 \mbox{Tr}
\left[P (\partial_1 P \partial_t P - \partial_t P \partial_1 P)\right]
\nonumber \\
&\in&\Pi_2[\CP(N-1)] = \Pi_2[SU(N)/U(N-1)] \nonumber \\
&=&\Pi_1[U(N-1)] = \Pi_1[U(1)] = \Z \,,
\end{eqnarray}
is the integer-valued topological charge. This term arises from Berry phases
associated with the underlying quantum spins which give rise to the vacuum
angle $\theta = n \pi$. Hence, the Berry phases cancel and yield $\theta = 0$ 
when $n$ is even, and they result in $\theta = \pi$ when $n$ is odd. As was
demonstrated in \cite{Jia11,Bea05}, for $N \geq 3$ the $\CP(N-1)$ model has a first
order phase transition at $\theta = \pi$, where the charge conjugation symmetry 
is spontaneously broken. In the rest of this paper, we will focus on
$\theta = 0$, i.e.\ even $n$, such that charge conjugation is not spontaneously
broken. Still, we will break charge conjugation explicitly by switching on 
chemical potentials.

Let us now consider how the chemical potential $\mu_a$ manifests itself
in the low-energy $\CP(N-1)$ model description. As a rule, chemical potentials
give rise to a Hermitean constant background vector potential that turns the
ordinary Euclidean time derivative $\partial_t P$ into a covariant derivative
\begin{equation}
D_t P = \partial_t P - [\mu_a T^a,P] \ .
\end{equation}
After dimensional reduction, the action of the $(1+1)$-d $\CP(N-1)$ model
thus takes the final form
\begin{eqnarray}
S[P] + i \theta Q[P]&=&\int_0^\beta dt \int_0^L dx_1 \frac{c}{g^2} 
\mbox{Tr}\left[\partial_1 P \partial_1 P + \frac{1}{c^2} D_t P D_t P\right]
\nonumber \\
&+&\frac{n}{2} \int_0^\beta dt \int_0^L dx_1 \mbox{Tr}
\left[P (\partial_1 P D_t P - D_t P \partial_1 P)\right] \ .
\end{eqnarray}

To summarize, we have presented an unconventional $(2+1)$-d $SU(N)$ 
antiferromagnetic quantum spin ladder regularization of the $(1+1)$-d $\CP(N-1)$
model, in which an odd extent $n = L'/a$ of the 2-direction gives rise to a 
non-trivial vacuum angle $\theta = \pi$. An external ``magnetic'' field applied
to the quantum spin ladder manifests itself as a chemical potential of the
emergent $\CP(N-1)$ model. This regularization makes the dynamics of $\CP(N-1)$ 
models accessible to quantum simulation experiments using ultracold 
alkaline-earth atoms in optical lattices \cite{Laf16,Laf15} . 

\section{The $SU(3)$ Spin Ladder and the $\CP(2)$ 
Model at non-zero Chemical Potential}

In this section, we investigate the $\CP(2)$ model at non-zero chemical 
potential using quantum Monte Carlo simulations (on a classical computer). This
can be used to validate future quantum simulation experiments of $SU(3)$ quantum
spin ladders using ultracold alkaline-earth atoms in an optical lattice.

\subsection{Phase Structure of the $SU(3)$ Quantum Spin 
Ladder}

\begin{figure}[tbh]
\begin{center}
\includegraphics[width=0.7\columnwidth]{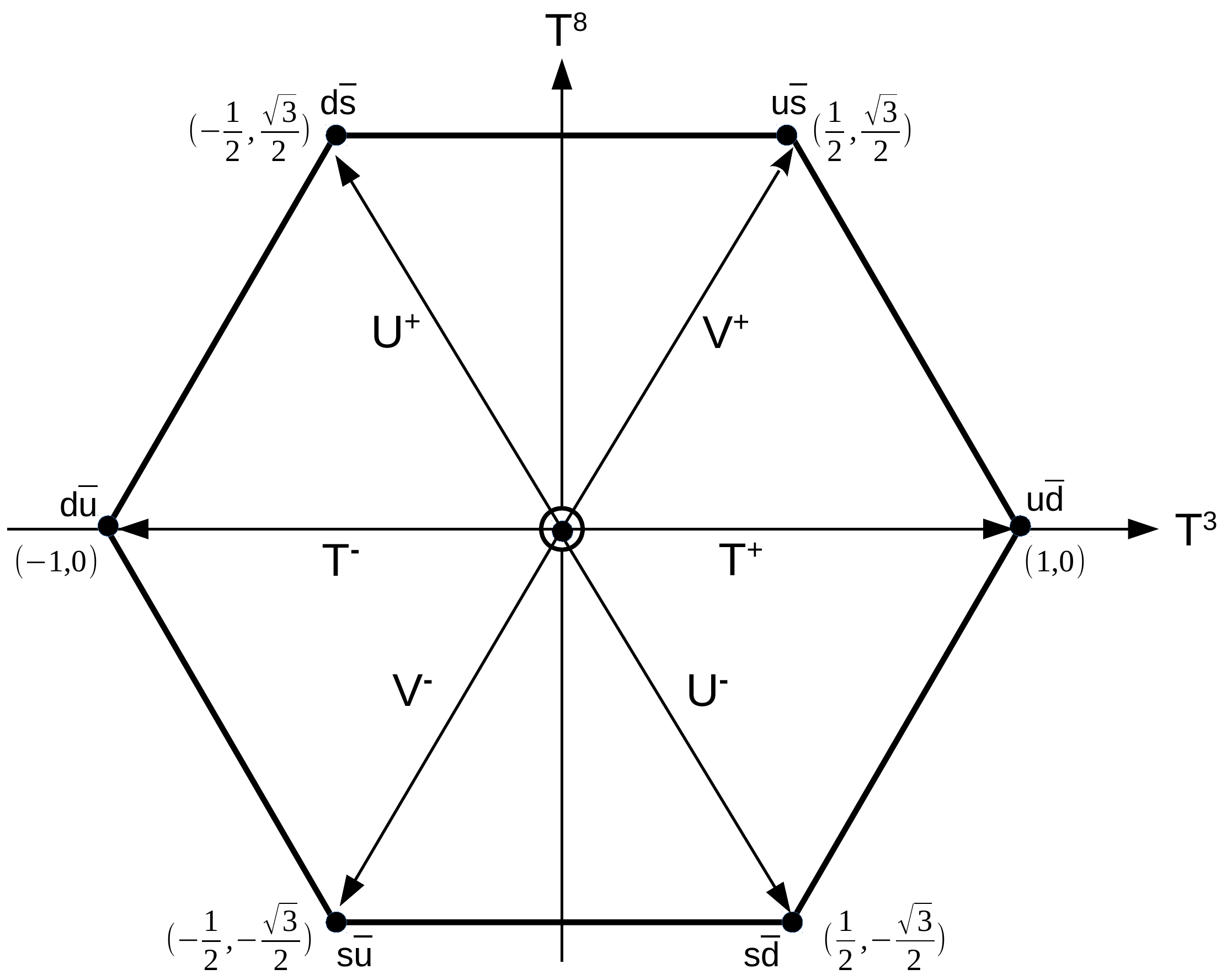}
\caption{\it $SU(3)$ octet of particles carrying the non-perturbatively
generated massgap $m$. The shift operators $T^\pm$, $U^\pm$, and $V^\pm$
(cf.\ eq.(\ref{shiftoperators})) can be used to generate the six states 
with non-zero $T^3$ and $T^8$ from the vacuum.}
\label{octet}
\end{center}
\end{figure}

\begin{figure}[tbh]
\begin{center}
\includegraphics[width=0.8\columnwidth]{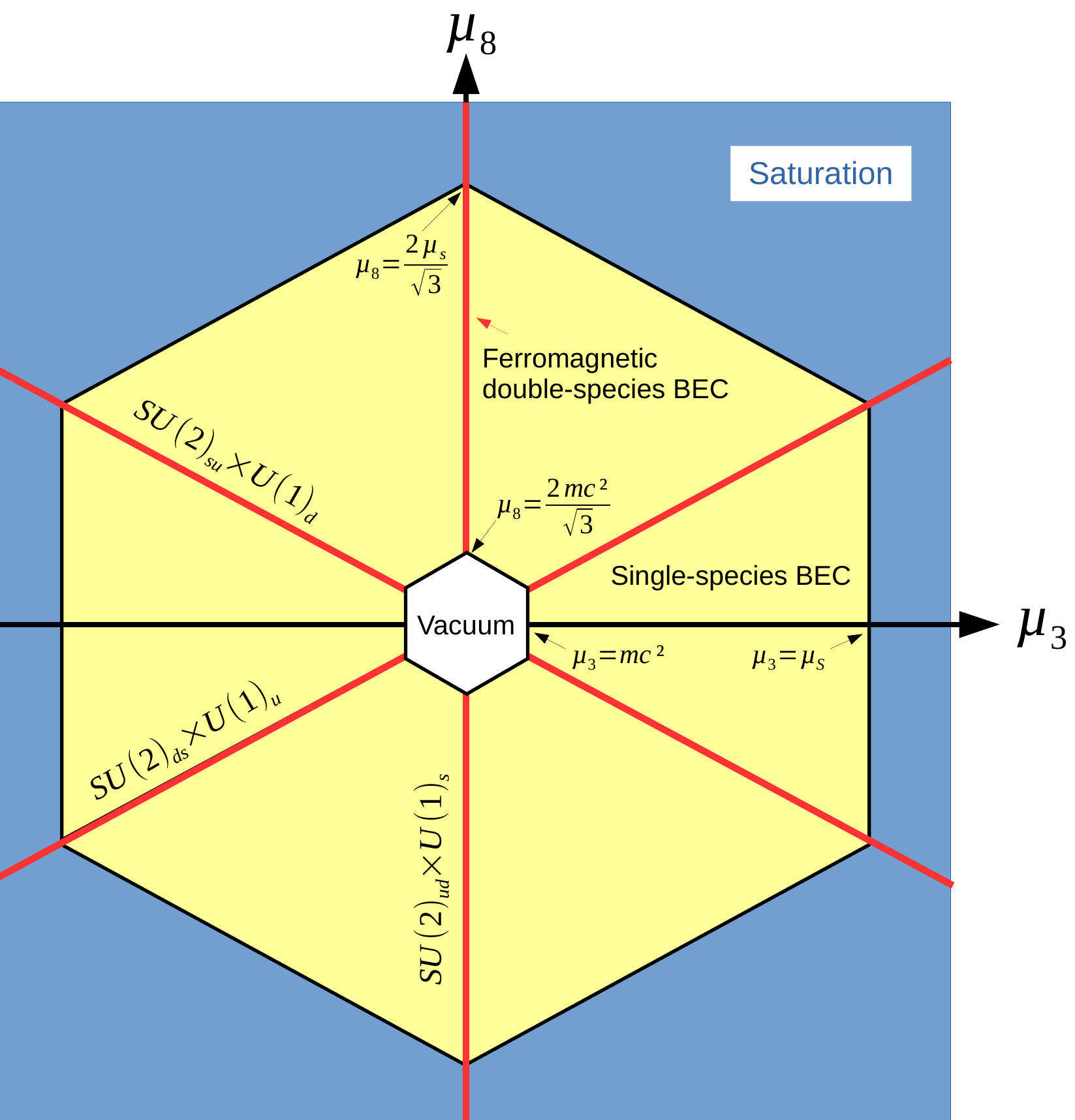}
\caption{\it Schematic phase diagram of the $SU(3)$ quantum spin ladder
in the $\mu_3,\mu_8$-plane at zero temperature. For small values
of the chemical potentials, the vacuum state exists in a hexagonal
region. On the other hand, for large values of the chemical potentials, 
outside of a large hexagon, the system saturates. In the region 
between the two hexagons, there exist different types of single-species
Bose-Einstein ``condensates'' (BEC). Generically, the chemical potentials 
break the global $SU(3)$ symmetry explicitly down to 
$U(1)_3 \times U(1)_8$. However, along the $\mu_8$-axis the symmetry
is enhanced to $SU(2)_{ud} \times U(1)_s$. Similarly, along the two 
lines at a 60 degree angle with respect to the $\mu_8$-axis, the 
symmetry is enhanced to $SU(2)_{ds} \times U(1)_u$ or
$SU(2)_{su} \times U(1)_d$. Along the lines of enhanced symmetry,
there exist different ferromagnetic double-species Bose-Einstein
``condensates''.}
\label{phasediagram}
\end{center}
\end{figure}

Let us consider possible phases of the $SU(3)$ quantum spin ladder at finite 
density. Here we restrict ourselves to even values of $n = L'/a$ which gives
rise to the $\CP(2)$ model at vacuum angle $\theta = 0$. 
Some results using periodic boundary conditions in the 2-direction were reported
in \cite{Eva16}. Here we use open boundary conditions which are more easily 
accessible in quantum simulation experiments. First of all, as a 
consequence of the Mermin-Wagner theorem, in $(1+1)$-d the global
$SU(3)$ symmetry cannot break spontaneously. After dimensional reduction the 
model has a massgap, since the lightest particles (which are no longer 
Goldstone bosons) pick up a mass $m$ non-perturbatively. As illustrated in 
Fig.\ref{octet}, these particles form an octet $\{8\}$ of $SU(3)$ with the
corresponding $T^3$ and $T^8$ ``charge'' assignments. When chemical potentials
$\mu_3$ and $\mu_8$ are switched on, some of these light particles are favored
over others. However, as long as the chemical potential is too small to overcome
the massgap, the system simply remains in the vacuum state, at least at zero
temperature. The particle with charges $(T^3,T^8) = (1,0)$ is favored over the 
vacuum if $\mu_3 > m c^2$. The particle with charges 
$(\frac{1}{2},\frac{\sqrt{3}}{2})$, on the other hand, is favored over the 
vacuum if $\frac{1}{2} \mu_3 + \frac{\sqrt{3}}{2} \mu_8 > m c^2$. As illustrated
in Fig.\ref{phasediagram}, there are six inequalities of this type, which 
define a regular hexagon around the origin, in which the vacuum state is 
favored. Along the three lines that connect the corners of the hexagon with the 
origin, the global $SU(3)$ symmetry is explicitly broken down to 
$SU(2) \times U(1)$ by the chemical potentials. For example, along the 
$\mu_8$-axis the symmetry is $SU(2)_{123}\times U(1)_8=SU(2)_{ud} \times U(1)_s$, while along the other 
two lines it is $SU(2)_{ds} \times U(1)_u$ or $SU(2)_{su} \times U(1)_d$. 
Everywhere else the chemical potentials break the $SU(3)$ symmetry explicitly 
to $U(1)_3 \times U(1)_8$.

As we will conclude from Monte Carlo simulations, once the chemical 
potentials overcome the massgap, depending on the segment of the phase diagram,
the bosons with quantum number combinations $u \overline{d}$, $u \overline{s}$,
$d \overline{s}$, $d \overline{u}$, $s \overline{u}$, or $s \overline{d}$ are 
produced in a second order phase transition. For example, once $\mu_3 > m c^2$
(while $\mu_8 = 0$) the particle density of the $u \overline{d}$ bosons 
increases continuously from zero. This is when the ``condensed matter physics''
of the $\CP(2)$ model sets in. It is then natural to ask what phase
of matter the bosons are forming. If different bosons would attract each other,
the system should phase separate. This is not what happens in this case. 
Instead, the bosons repel each other and form a gas. We will
present numerical evidence that (at least at zero temperature) the ultracold
bosonic gas ``condenses''. More precisely, the $U(1)_3$ symmetry undergoes the 
Kosterlitz-Thouless phenomenon. In view of the Mermin-Wagner theorem, this is 
as close as a $(1+1)$-d system can come to Bose-Einstein condensation. 

In other segments of the phase diagram, bosons with other quantum number 
combinations ``condense'', and corresponding linear combinations of $U(1)_3$ 
and $U(1)_8$ undergo the Kosterlitz-Thouless phenomenon. These Bose-Einstein 
``condensates'' consist of a single species of bosons with one of the six 
quantum number combinations listed above. While for 
$\mu_3 > 0$ and $|\mu_8| < \frac{1}{\sqrt{3}} \mu_3$ the bosons of type  
$u \overline{d}$ ``condense'', such that the $U(1)_3 = U(1)_{ud}$ (but not the 
$U(1)_8 = U(1)_s$) symmetry is affected by the Kosterlitz-Thouless phenomenon, 
for $\mu_3 > 0$ and $\mu_8 > \frac{1}{\sqrt{3}} \mu_3$, the bosons 
of type $u \overline{s}$ ``condense'', i.e.\ the $U(1)_{us}$ (but not the 
$U(1)_d$) symmetry is affected by the Kosterlitz-Thouless phenomenon. Note that
$U(1)_{su}$ and $U(1)_d$ are generated by
\begin{equation}
\frac{1}{2}(T^3 + \sqrt{3} T^8) = \frac{1}{2} \mbox{diag}(1,0,-1), \quad
\frac{1}{2}(\sqrt{3} T^3 - T^8) = \frac{1}{2 \sqrt{3}} \mbox{diag}(1,-2,1),
\end{equation}
respectively. Analogously, $U(1)_{ds}$ and $U(1)_u$ are generated by
\begin{equation}
- \frac{1}{2}(T^3 - \sqrt{3} T^8) = \frac{1}{2} \mbox{diag}(0,1,-1), \quad
- \frac{1}{2}(\sqrt{3} T^3 + T^8) = \frac{1}{2 \sqrt{3}} \mbox{diag}(-2,1,1).
\end{equation}
Indeed, for $\mu_3 > 0$ and $\mu_8 < - \frac{1}{\sqrt{3}} \mu_3$, the bosons 
of type $s \overline{d}$ ``condense'', and now the $U(1)_{ds}$ (but not the 
$U(1)_u$) symmetry is affected by the Kosterlitz-Thouless phenomenon. The 
situation is analogous for $\mu_3 < 0$.

A special situation arises when the $U(1)_3 \times U(1)_8$ symmetry is enhanced
to a non-Abelian symmetry. For example, when $\mu_3 = 0$ the $U(1)_3$ symmetry 
is enhanced to $SU(2)_{123}=SU(2)_{ud}$. In that case, two species of bosons --- namely 
those of the types $u \overline{s}$ and $d \overline{s}$ --- ``condense''. As a
consequence, now the $U(1)_8 = U(1)_s$ symmetry is affected by the 
Kosterlitz-Thouless phenomenon. How does the corresponding double-species 
Bose-Einstein ``condensate'' realize the enhanced $SU(2)_{ud}$ symmetry? 
Interestingly, our Monte Carlo simulations demonstrate that the 
system is a ferromagnet with a conserved order parameter --- the
$SU(2)_{ud}$ vector $(T^1,T^2,T^3)$. It should be stressed that the
Mermin-Wagner theorem does not prevent ferromagnetism in $(1+1)$ dimensions.

At very large values of the chemical potentials, the system ultimately 
saturates. In particular, for $\mu_3 > \mu_s$ and 
$|\mu_8| < \frac{1}{\sqrt{3}} \mu_3$ at zero temperature the spins on sublattice
$A$ are in the state $u$ while the spins on sublattice $B$ are in the state 
$\overline{d}$. It is straightforward to determine the 
exact value of $\mu_s \propto J$ associated with saturation.
We will present this calculation elsewhere. 
Similarly, for $\frac{1}{2} \mu_3 + \frac{\sqrt{3}}{2} \mu_8 > \mu_s$, 
$\mu_3 > 0$, and $\mu_8 > \frac{1}{\sqrt{3}} \mu_3$ the spins on sublattice
$A$ are still in the state $u$ while the spins on sublattice $B$ are now in the 
state $\overline{s}$. As a consequence, for large values of the chemical
potentials there is a region (bounded by a large hexagon) in which the system
saturates in one of the six states $u \overline{d}$, $u \overline{s}$,  
$d \overline{s}$, $d \overline{u}$, $s \overline{u}$, and $s \overline{d}$. 

It is interesting to ask whether there are other additional phases at 
intermediate values of the chemical potentials, before one reaches saturation. 
We will address this question in the future. In this paper, we concentrate on
moderate values of the chemical potentials and we establish the
existence of two non-trivial phases --- the single-species Bose-Einstein 
``condensate'' for $\mu_3 > m c^2$, $\mu_8 = 0$, and the double-species 
ferromagnet for $\mu_3 = 0$, $\mu_8 > \frac{2}{\sqrt{3}} m c^2$.

The previous discussion of the phase diagram referred to strictly zero
temperature. Then the $(1+1)$-d quantum system experiences the 
Kosterlitz-Thouless phenomenon, i.e.\ bosons ``condense'' and the correlation
length (of the infinite system with both $L \rightarrow \infty$ and inverse 
temperature $\beta \rightarrow \infty$) diverges. Even at an infinitesimally 
small non-zero temperature, the correlation length becomes finite and the 
previously discussed phase transitions (as a function of $\mu_3$ and $\mu_8$ 
at zero temperature) are washed out to smooth cross-overs.

\subsection{Spinwave Velocity}

First of all, it is useful to determine the low-energy parameters of the
$(2+1)$-d $\CP(2)$ effective field theory before dimensional reduction.
These are the spin stiffness $\rho_s$ and the spinwave velocity $c$. Here
we determine the value of $c$ using the method described in \cite{Jia11},
which results in $c = 1.7763(2) J a$ (where $J$ is the exchange coupling
and $a$ is the lattice spacing). In the continuum limit (which is 
approached by gradually increasing $L'$) the $(1+1)$-d $\CP(2)$ model 
results as a relativistic quantum field theory via dimensional reduction. 
As a consequence of the emerging symmetry between space and Euclidean
time, the correlation lengths $\xi$ and $\xi_t$ in space and time should
be related by $\xi = c \xi_t$. By measuring the exponential decay of 
the 2-point function $\langle T_x^a T_y^a \rangle$ at vanishing chemical 
potentials ($\mu_3 = \mu_8 = 0$), we have explicitly verified this 
relation. 

\subsection{Single-Particle States}

The mass gap separates the vacuum from an $SU(3)$ octet of 
massive particles with a rest energy $m c^2 = 1/\xi_t$ (or equivalently 
a rest mass $m = 1/(c \xi)$). For $L'/a = 10, 12, 14$ and $L/a = 250$ 
we obtain the rest energy values $m c^2 = 0.1006(1) J$, $0.05873(1) J$, 
and $0.0344(1) J$, respectively, which correspond to spatial correlation 
lengths (or equivalently Compton wave lengths) $\xi = 17.66(2) a$, 
$30.25(1) a$, and $51.6(2) a$. This confirms the exponential
increase of $\xi$ (cf.\ eq.(\ref{xi})) which is a signature of 
asymptotic freedom. Note that $\xi \ll L$, such that finite-size 
effects can be neglected.

\begin{figure}[tbh]
\begin{center}
\includegraphics[width=\columnwidth]{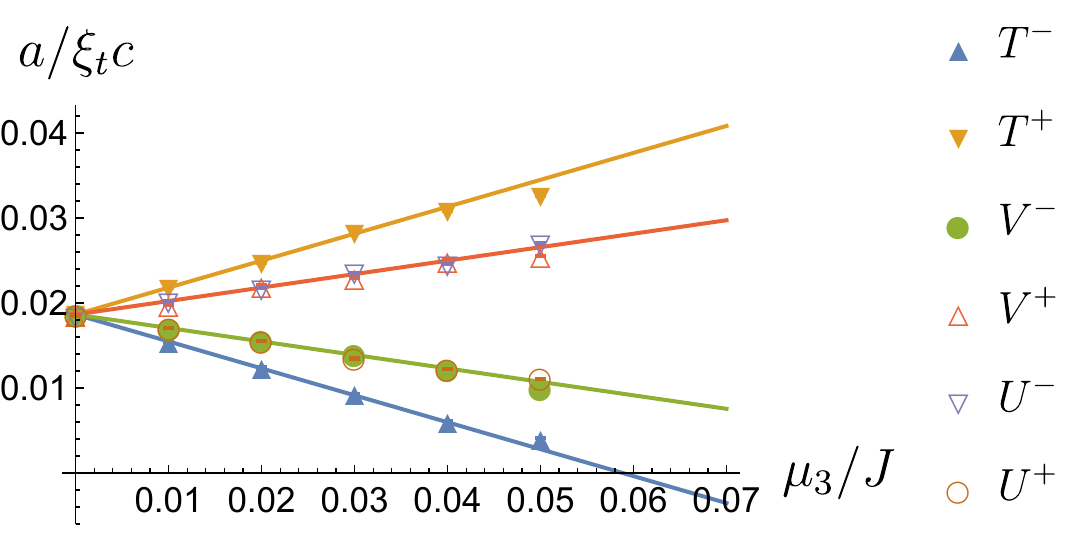} \\
\includegraphics[width=\columnwidth]{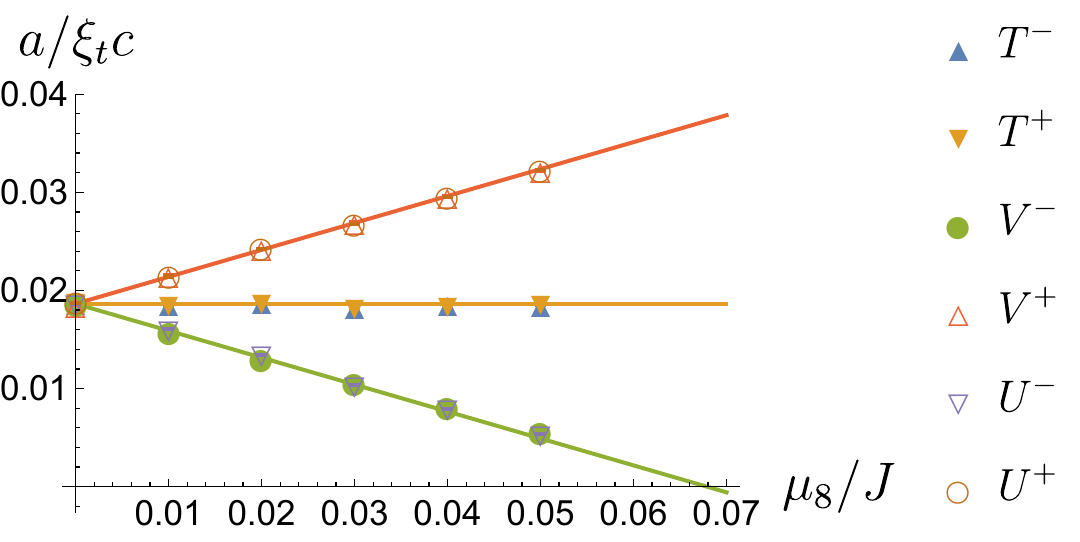}
\caption{\it Inverse correlation lengths as a function of $\mu_3$,
putting $\mu_8 = 0$ (top), and as a function of $\mu_8$, putting
$\mu_3 = 0$ (bottom). The Monte Carlo data confirm that the
eigenvalues of $H - \mu_a T^a$ are linear functions of the chemical
potentials. The slopes of the straight lines (which are analytic 
predictions and not fits) are determined by the charges $(T^3,T^8)$ 
of the corresponding single-particle states.}
\label{massgap}
\end{center}
\end{figure}

In order to verify that the massive single-particle states indeed
form an $SU(3)$ octet, we have analyzed the exponential decay of the
Euclidean-time correlation functions of the various states both as a 
function of $\mu_3$ and $\mu_8$. The eigenvalues of $H - \mu_a T^a$
for the various single-particle states (at zero spatial momentum) are 
given by $m c^2 - \mu_3 T^3 - \mu_8 T^8$, and are thus linear functions 
of the chemical potentials. Fig.\ref{massgap} shows the corresponding
inverse correlation lengths extracted from correlation functions
associated with the $SU(3)$ shift operators $T^\pm$, $U^\pm$, and
$V^\pm$ (cf.\ eq.(\ref{shiftoperators})) as functions of $\mu_3$ 
(putting $\mu_8 = 0$) and as functions of $\mu_8$ (putting 
$\mu_3 = 0$). Acting on the vacuum, the various shift operators 
generate six single-particle states that are members of an $SU(3)$ 
octet with the quantum number combinations $(T^3,T^8) = (\pm 1,0)$, 
$(\pm \frac{1}{2},\pm \frac{\sqrt{3}}{2})$. Indeed, these quantum 
numbers are verified by the observed linear $\mu_3$- and 
$\mu_8$-dependencies of the corresponding inverse correlation lengths.

\subsection{Leaving the Vacuum: the Onset of Particle 
Production}

In this subsection we investigate the critical values of the chemical 
potentials that separate the vacuum sector of the phase diagram from the
particle sectors in which the ``condensed matter physics'' of the $\CP(2)$
model takes place. We will address the nature of the phase formed by these
particles (the Bose-Einstein ``condensates'' mentioned above) in the next
subsection.

\begin{figure}[tb]
\begin{center}
\includegraphics[width=\columnwidth]{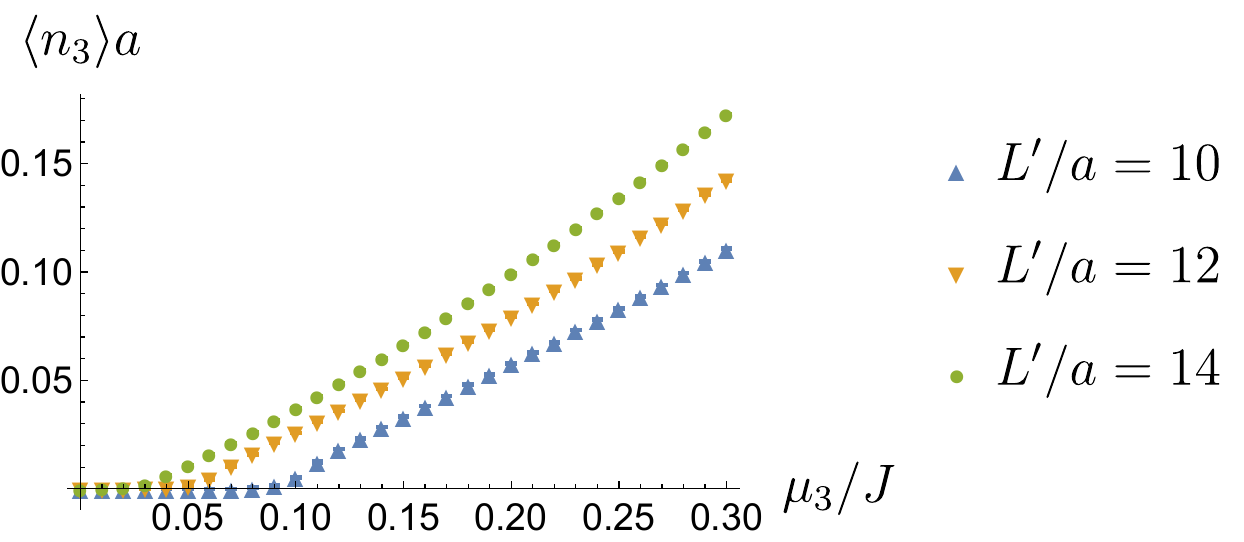} \\
\includegraphics[width=\columnwidth]{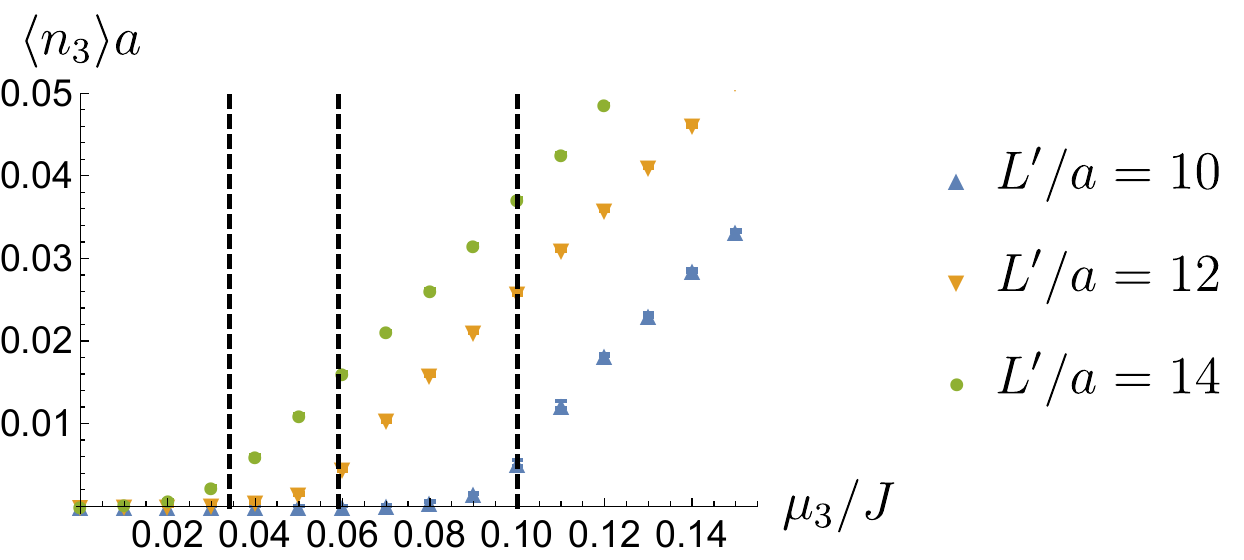}
\caption{\it Particle density $\langle n_3\rangle$ as a function of
$\mu_3$ (putting $\mu_8 = 0$) for three different values of $L'$
over a wide range of $\mu_3$ (top) and zooming into a narrower
range around the onsets (bottom), keeping $\beta c = L = 250 a$
fixed. The critical values $\mu_3 = m c^2$ for the onset of particle 
production at zero temperature are indicated by the three dashed 
lines, corresponding to $L'/a = 10, 12, 14$.}
\label{n3onset}
\end{center}
\end{figure}

Let us consider the effect of a chemical potential $\mu_3 > 0$, 
$\mu_8 = 0$ on the expectation value 
\begin{equation}
\langle n_3\rangle = \frac{1}{L} \langle T^3\rangle,
\end{equation}
which plays the role of the density of the particles of type 
$u \overline{d}$ (cf.\ Fig.\ref{n3onset}). Indeed, as indicated by the 
dashed lines, we see that the onset of particle production occurs near 
$\mu_3 = m c^2$. At zero temperature, no particles should be produced 
below this critical value. For finite temperature, on the other hand, 
particles can also be produced by thermal fluctuations, thus washing 
out the onset behavior. Here the inverse temperature was fixed to 
$\beta J = 140.75$ and again $L/a = 250$. This corresponds to a
square-shaped Euclidean space-time volume with $L = \beta c$.

\begin{figure}[tb]
\begin{center}
\includegraphics[width=\columnwidth]{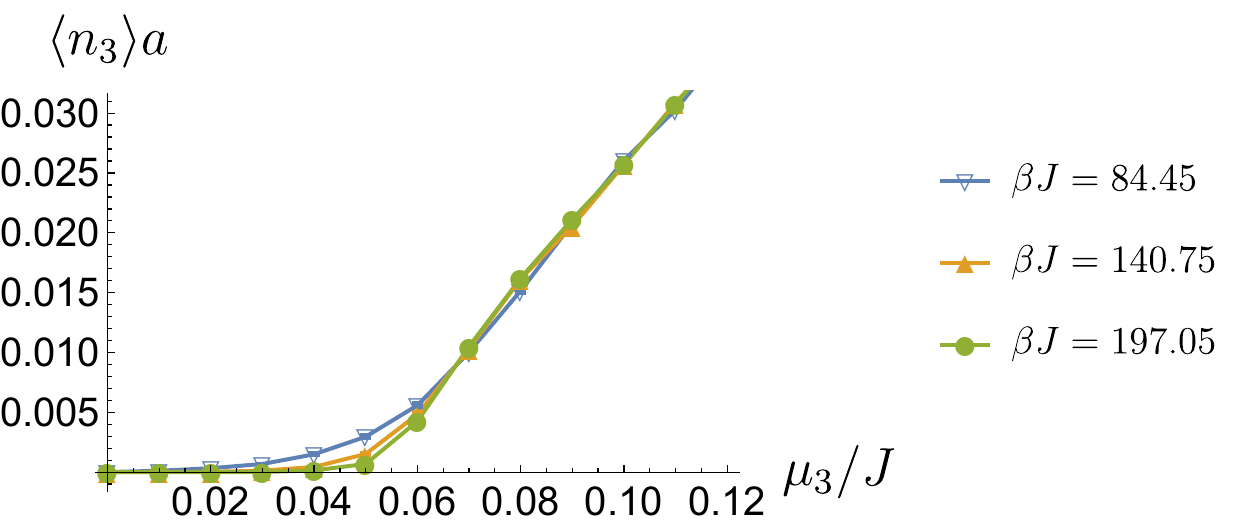} \\
\includegraphics[width=\columnwidth]{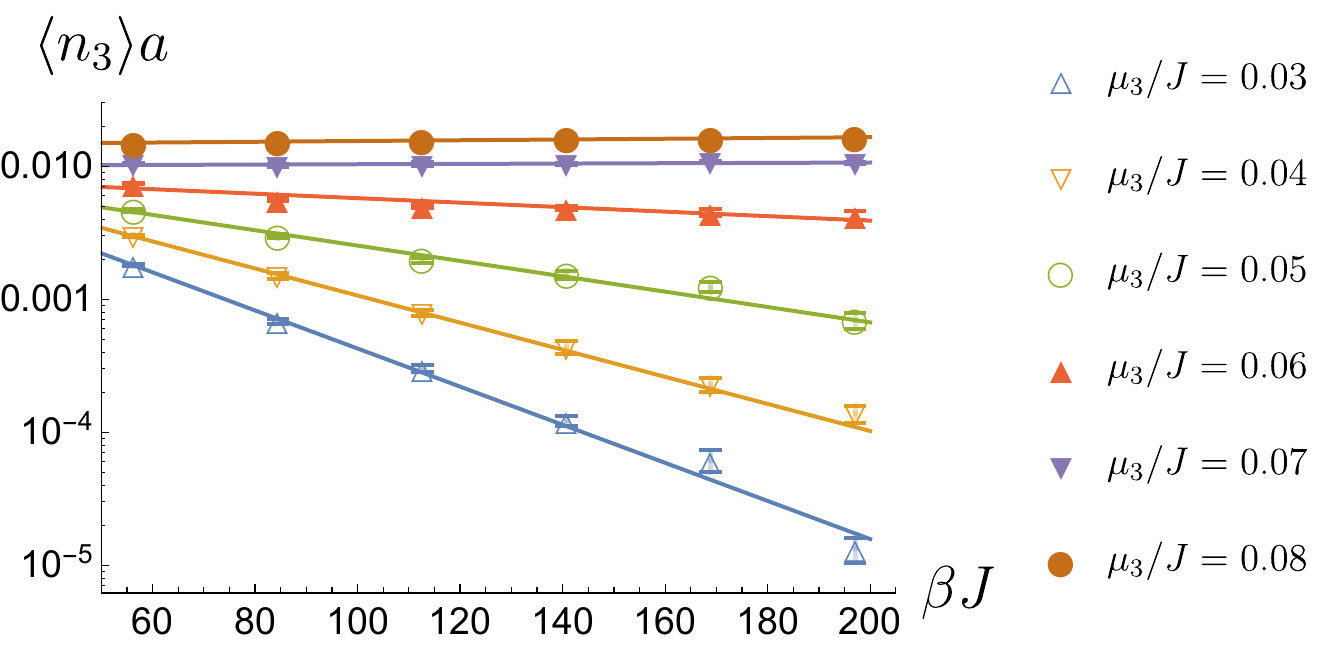}
\caption{\it Particle density $\langle n_3\rangle$ as a function of
$\mu_3$ (putting $\mu_8 = 0$) for three different inverse temperatures
$\beta$, with lines to guide the eye (top), and as a function of 
$\beta$ for six different values of $\mu_3$ with exponential fits
(bottom). Below the critical chemical potential $\mu_3 = m c^2$, the 
particle density vanishes in the zero-temperature limit.}
\label{n3mu3}
\end{center}
\end{figure}

For the rest of the paper we set $L'/a = 12$, which puts us 
comfortably close to the continuum limit. In order to investigate 
finite-temperature and finite-volume effects in the onset region, 
in Fig.\ref{n3mu3} we compare $\langle n_3\rangle$ for three values of 
the inverse temperature $\beta J = 56.3$, $140.75$, and $197.05$.
Indeed, as long as $\mu_3 < m c^2$, the particle density goes to zero
with decreasing temperature, while for $\mu_3 > m c^2$ it approaches
a non-zero value. This is confirmed by logarithmically plotting the 
same data as a function of $\beta$ now keeping $\mu_3$ fixed (cf.\
Fig.\ref{n3mu3}). In particular, we now see that the particle 
density $\langle n_3 \rangle$ decreases exponentially with $\beta$ 
for $\mu_3 < m c^2$.

\begin{figure}[tb]
\begin{center}
\includegraphics[width=\columnwidth]{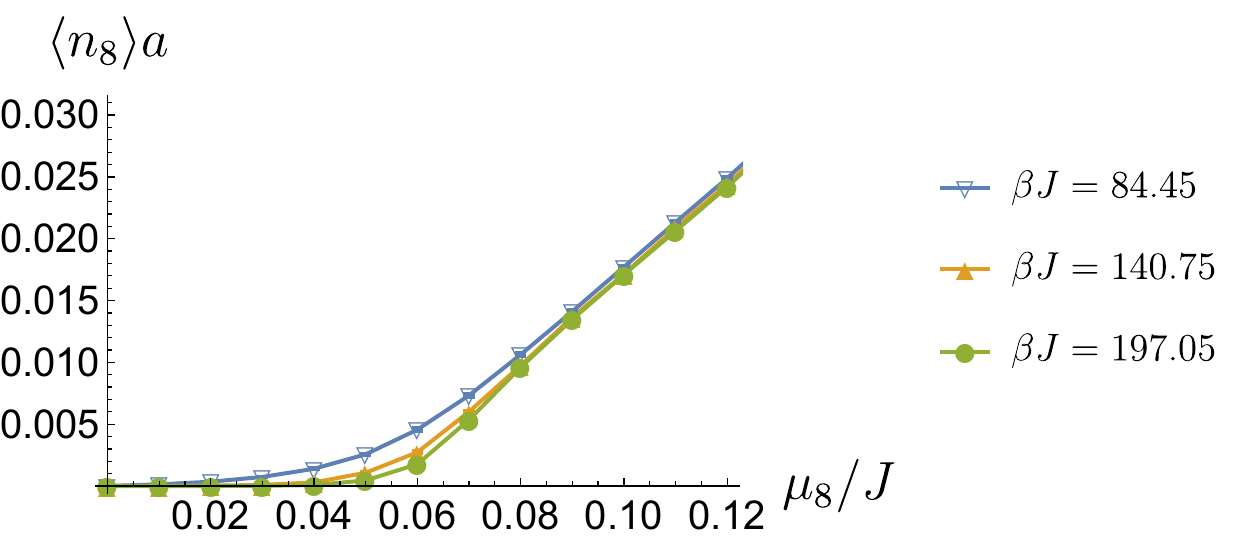} \\
\includegraphics[width=\columnwidth]{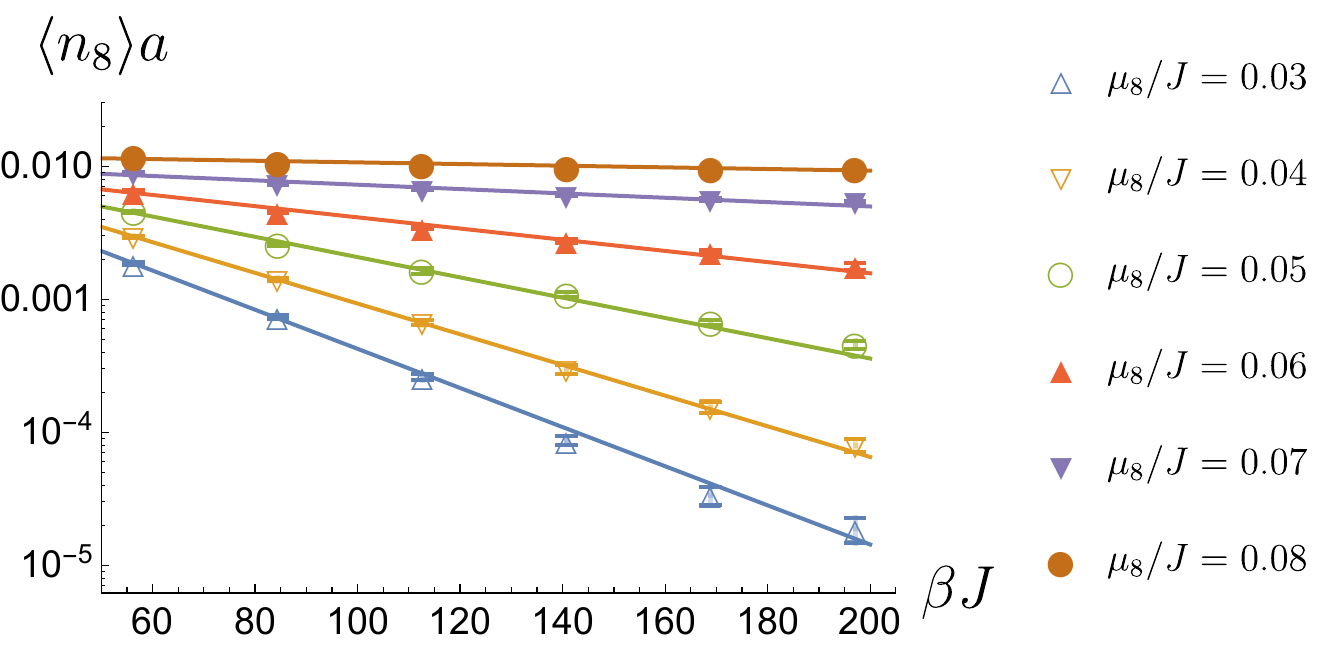}
\caption{\it Particle density $\langle n_8\rangle$ as a function of
$\mu_8$ (putting $\mu_3 = 0$) for three different inverse temperatures
$\beta$, with lines to guide the eye (top), and as a function of 
$\beta$ for six different values of $\mu_8$ with exponential fits
(bottom). Below the critical chemical potential 
$\mu_8 = 2 m c^2/\sqrt{3}$, the particle density vanishes in the 
zero-temperature limit.}
\label{n8mu8}
\end{center}
\end{figure}

Let us now consider the situation with $\mu_3 = 0$ and $\mu_8 > 0$. 
Then the chemical potentials break the $SU(3)$ symmetry explicitly 
down to $SU(2)_{ud} \times U(1)_s$. In particular, now the particles 
of type $u \overline{s}$ and $d \overline{s}$ are both equally 
favored by the chemical potential $\mu_8$. The onset of the density
\begin{equation}
\langle n_8\rangle = \frac{1}{L} \langle T^8\rangle,
\end{equation}
again for $\beta J = 56.3$, $140.75$, and $197.05$, is illustrated 
in Fig.\ref{n8mu8}. Note that the critical value for particle
production is now given by $\mu_8 = \frac{2}{\sqrt{3}} m c^2$, 
because the particles of type $u \overline{s}$ and $d \overline{s}$ 
have $T^8 = \frac{\sqrt{3}}{2}$. As before, the data are replotted 
logarithmically as a function of $\beta$, which again shows an 
exponential suppression of the particle density $\langle n_8 \rangle$ 
with temperature below the onset.

\begin{figure}[tb]
\begin{center}
\includegraphics[width=\columnwidth]{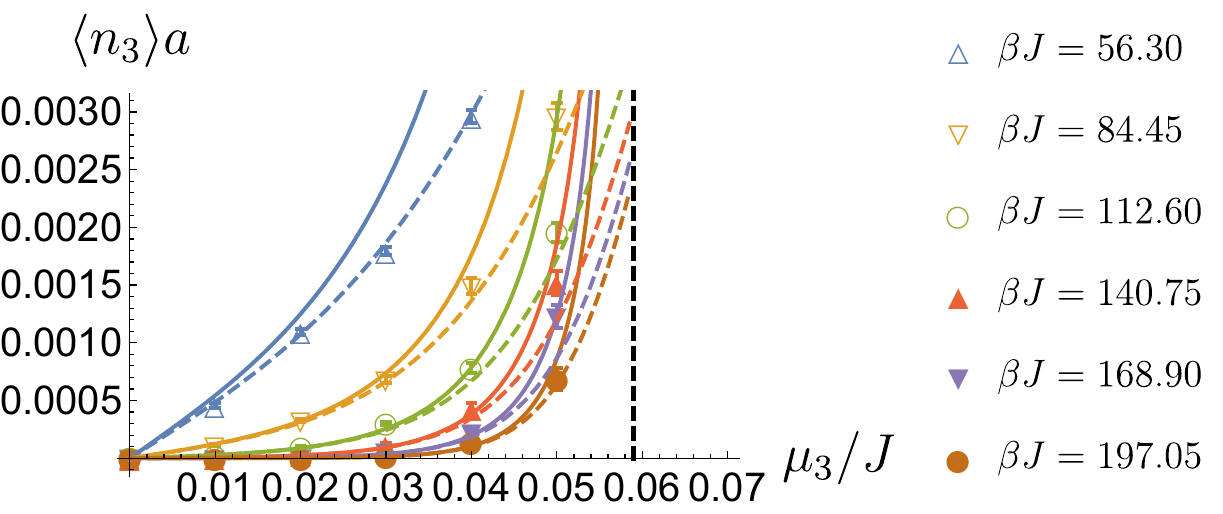} \\
\includegraphics[width=\columnwidth]{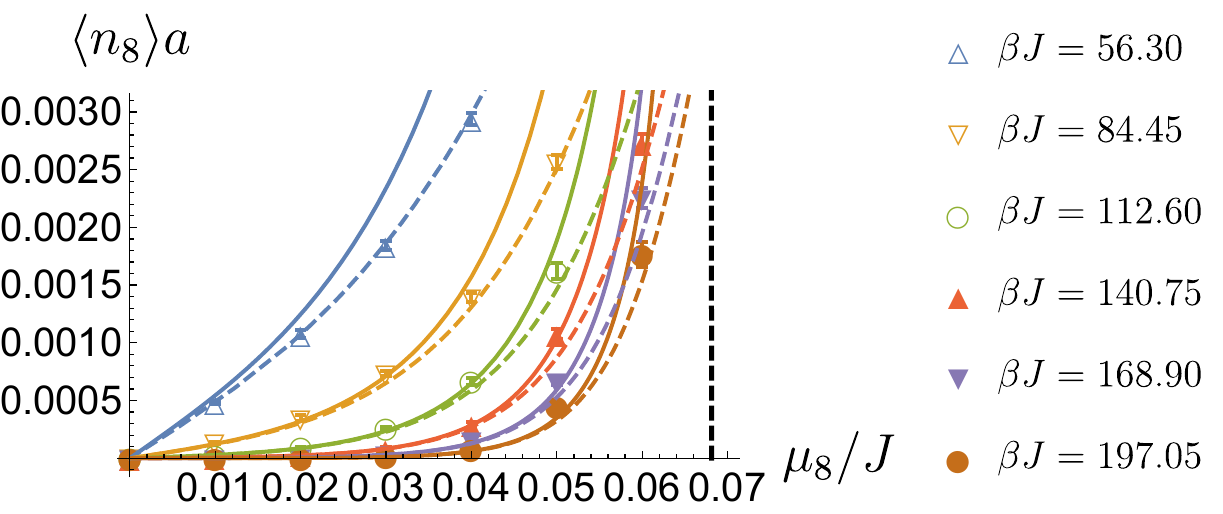}
\caption{\it Particle density $\langle n_3\rangle$ as a function of
$\mu_3$, putting $\mu_8 = 0$ (top), and $\langle n_8\rangle$ as a 
function of $\mu_8$, putting $\mu_3 = 0$ (bottom), for six different 
inverse temperatures $\beta$. Monte Carlo data are compared with a
model of free bosons (solid lines) and a model of free fermions
(dashed lines). The Monte Carlo data lie between the predictions
of the bosonic and the fermionic model, indicating that the bosons 
of the $\CP(2)$ model are not free but repel each other.}
\label{n3n8model}
\end{center}
\end{figure}

In Fig.\ref{n3n8model} we compare the Monte Carlo data for 
$\langle n_3 \rangle$ and $\langle n_8 \rangle$ 
(at $\beta J = 197.05$ and $L/a = 300$) with analytic
predictions of a free boson and a free fermion model. Although the 
massive particles in the $(1+1)$-d $\CP(2)$ model are bosons, they 
are not free but are expected to repel each other. In $(1+1)$-d, 
bosons with an infinite short-range repulsion are equivalent to free 
fermions. Indeed, the Monte Carlo data lie between the predictions 
of the free boson and the free fermion model, thus indicating that 
the $\CP(2)$ bosons have a finite repulsive interaction. Below the
critical values of the chemical potentials, the particle density
is very low. Consequently, both the bosonic and the fermionic model
agree with each other. The fact that the Monte Carlo data of the
$\CP(2)$ model are correctly represented in this regime (without 
fitting any parameters), confirms that the bosons of the $\CP(2)$ 
model indeed have the mass $m$.

The free boson model assumes that we have particles and 
anti-particles of mass $m$ with a relativistic dispersion relation 
$E(p) = \sqrt{(m c^2)^2 + (p c)^2}$ in a finite spatial interval of 
size $L$ with periodic boundary conditions. Hence, the momenta $p$ 
are quantized in integer multiples of $2 \pi/L$. The particle 
densities are given by
\begin{equation}
\langle n_3 \rangle = 
\frac{\partial \log Z(\beta,\mu_3,\mu_8)}{\partial (\beta \mu_3)},
\quad \langle n_8 \rangle = 
\frac{\partial \log Z(\beta,\mu_3,\mu_8)}{\partial (\beta \mu_8)}.
\end{equation}
The grand canonical partition function factorizes in momentum and 
$T^3$ and $T^8$ quantum number sectors
\begin{equation}
Z(\beta,\mu_3,\mu_8) = \prod_{p,T^3,T^8} Z_{p,T^3,T^8}(\beta,\mu_3,\mu_8),
\end{equation}
with
\begin{equation}
Z_{p,T^3,T^8}(\beta,\mu_3,\mu_8) = 
\sum_n \exp(- \beta (E(p) - \mu_3 T^3 - \mu_8 T^8) n).
\end{equation}
For free bosons, the occupation numbers of each mode take values
$n \in \{0,1,2,\dots,\infty\}$. In order to mimic boson repulsion,
we also introduce a ``fermionic'' model, simply by restricting the
occupation numbers to $n \in \{0,1\}$. One then obtains
\begin{eqnarray}
\langle n_3 \rangle&=&\frac{1}{L} \sum_{p,T^3,T^8}
\frac{T^3}{\exp(\beta(E(p) - \mu_3 T^3 - \mu_8 T^8)) \pm 1}, 
\nonumber \\
\langle n_8 \rangle&=&\frac{1}{L} \sum_{p,T^3,T^8}
\frac{T^8}{\exp(\beta(E(p) - \mu_3 T^3 - \mu_8 T^8)) \pm 1} \ .
\end{eqnarray}
Here $\pm$ corresponds to fermions and bosons, respectively. Note
that free bosons give rise to an infinite density once the chemical
potentials reach their critical values. The sum extends over all 
momenta $p = 2 \pi l/L$ (with $l \in \Z$) and over the 8 particle 
and anti-particle states in the $SU(3)$ octet. The 2 states with 
$T^3 = T^8 = 0$ do not contribute, because the corresponding 
particles are neutral. 

It should be noted that the particles of the ``fermionic'' model
are not truly relativistic two-component Dirac fermions, but rather 
one-component objects with fermionic statistics. This violates the 
spin-statistics theorem and is thus inconsistent with relativistic 
invariance. In any case, neither the results of the bosonic nor of 
the ``fermionic'' model are expected to describe the behavior of the 
$\CP(2)$ model completely correctly. They just indicate that the 
bosons of the $\CP(2)$ model repel each other. It would also be interesting
to compare the Monte Carlo data with the non-relativistic Lieb-Liniger
model \cite{Lie60}. This will be addressed elsewhere.

\subsection{Single-Species Bose-Einstein ``Condensation''}

In this subsection, we put $\mu_8 = 0$ and study the physics as a 
function of $\mu_3$. We have investigated systems of two different 
sizes $L = 300 a$ and $600 a$ (with fixed $L' = 12 a$) at two 
different inverse temperatures $\beta J = 197.05$ and $281.5$. 

\begin{figure}[tbh]
\begin{center}
\includegraphics[width=0.52\columnwidth]
{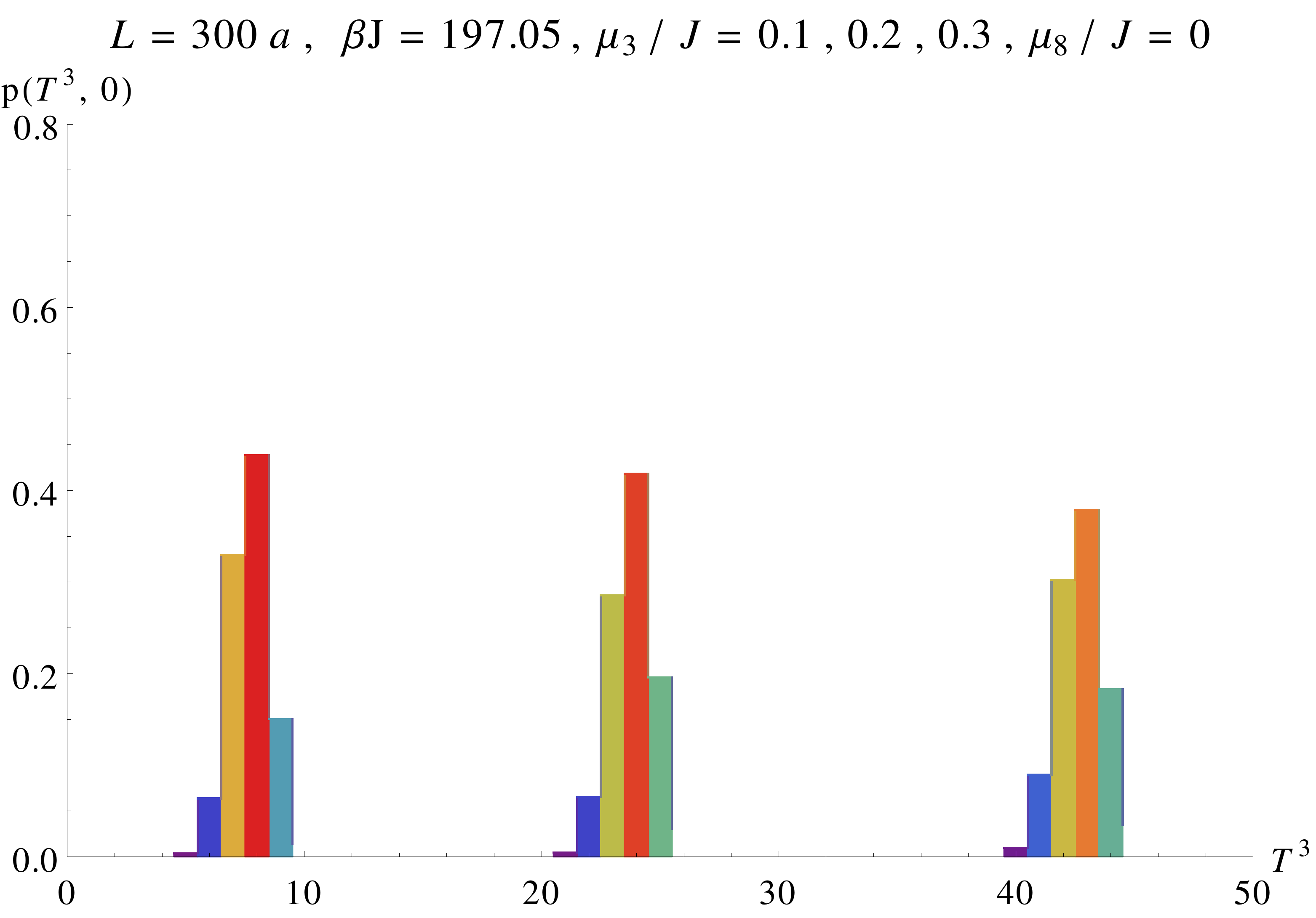} \\
\includegraphics[width=0.52\columnwidth]
{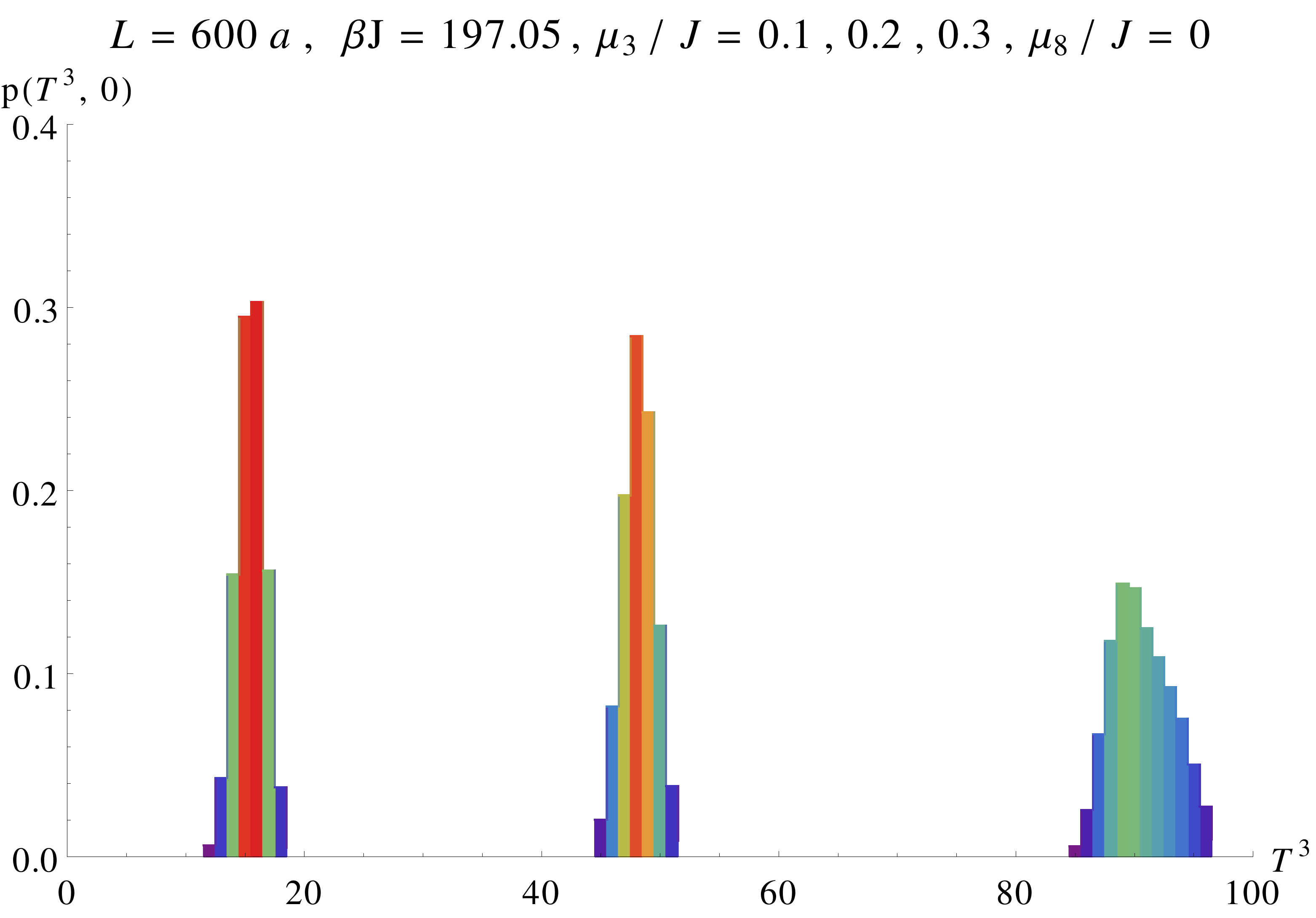} \\
\includegraphics[width=0.52\columnwidth]
{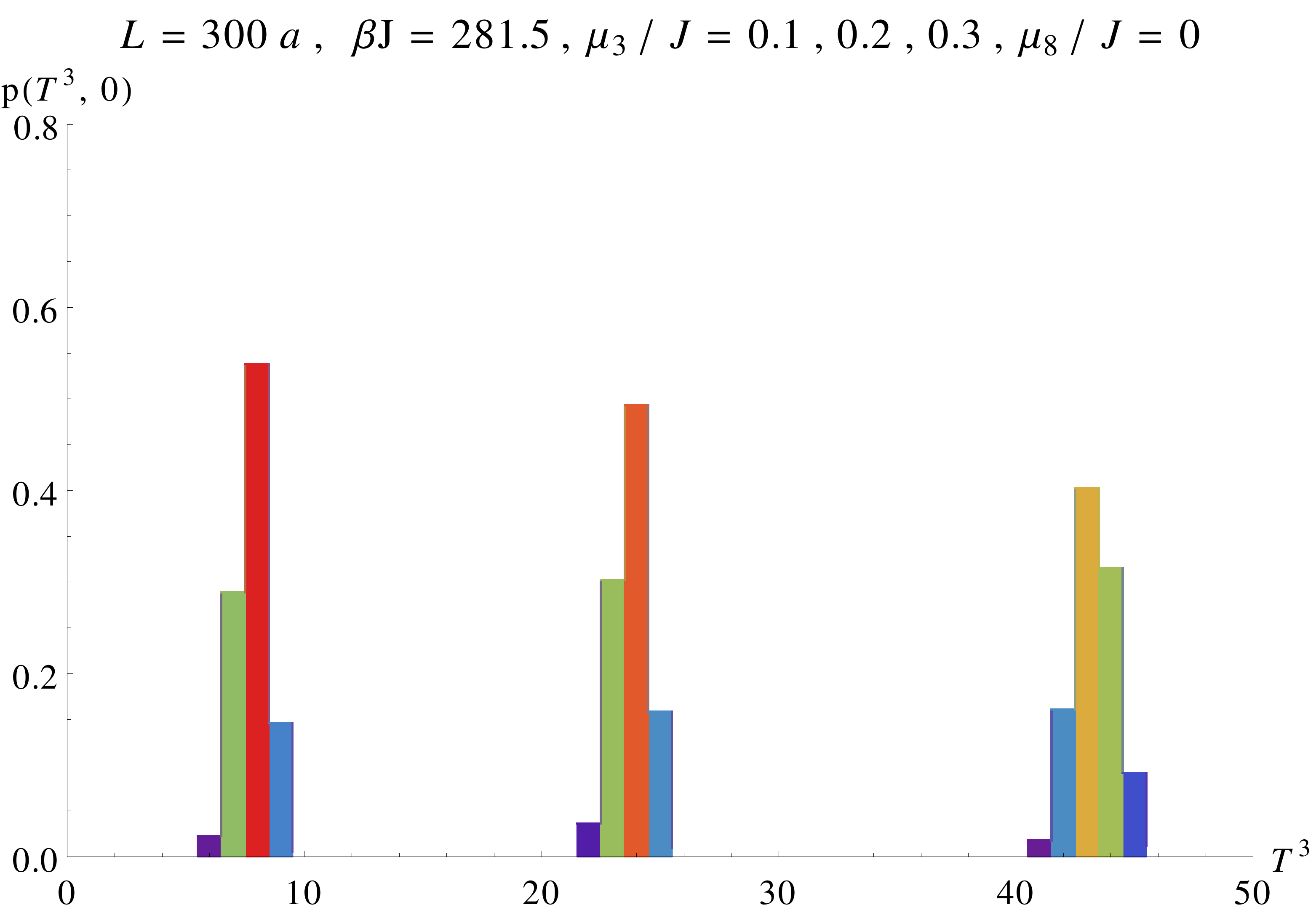}
\caption{\it Probability distributions $p(T^3,T^8 = 0)$ for three 
different values of the chemical potential $\mu_3 = 0.1, 0.2, 0.3$
and $\mu_8 = 0$. Lattice size and inverse temperature are fixed to
$L = 300 a$, $\beta J = 197.05$ (top), $L = 600 a$, 
$\beta J = 197.05$ (middle), and $L = 300 a$, $\beta J = 281.5$ 
(bottom).}
\label{histT3}
\end{center}
\end{figure}

The probability distributions of the different charge sectors $(T^3,T^8)$
are illustrated in Fig.\ref{histT3} (top) for $L = 300 a$ and 
$\beta J = 197.05$ for $\mu_3/J = 0.1, 0.2, 0.3$, and $\mu_8 = 0$. For 
$\mu_3 = 0.1 J > m c^2$, we see the onset of particle production with
$\langle T^3 \rangle > 0$. As expected for $\mu_8 = 0$, almost all 
states have $T^8 = 0$, which implies that the produced bosons carry
the quantum number combination $u \overline{d}$. For $\mu_3 = 0.1 J$,
there are typically 8 bosons in the $L = 300 a$ box. Since the 
Compton wave length of the bosons is $\xi = 1/(m c) = 30.25 a$, they
are not very dilute at this value of $\mu_3$. When the chemical 
potential is increased further, the probability distribution is shifted 
to larger values of $T^3$. For $\mu = 0.2 J$ the most probable particle 
number is 24, and for $\mu_3 = 0.3 J$ it is 43, which correspond to
rather dense systems of bosons. Fig.\ref{histT3} (middle) shows the 
same situation for the larger spatial volume $L = 600 a$. Now, for a 
given value of $\mu_3$, about twice as many particles are being produced, 
but their density $\langle T^3\rangle/L$ remains more or less the same, 
indicating that finite-size effects are moderate. Fig.\ref{histT3} 
(bottom) shows results for the $L = 300 a$ box at the lower temperature 
corresponding to $\beta J = 281.5$. As expected, thermal fluctuations 
in the charge (or equivalently particle number) distribution are then 
further suppressed. From these results we conclude that, as $\mu_3$ 
increases from $m c^2$ to $0.3 J$, the system contains an increasing 
density of bosons of type $u \overline{d}$.

It is natural to ask what state of ``$\CP(2)$ condensed matter'' these
bosons are forming. As we will now demonstrate, not surprisingly, in 
the zero-temperature limit (given the limitations of the Mermin-Wagner
theorem), they form a Bose-Einstein ``condensate''. To study this, we
have investigated the spatial winding numbers, $W_3, W_8 \in \Z$, for
which 
\begin{eqnarray}
\langle W_3^2 \rangle&=&
\frac{\partial^2 \log Z(\beta,L,L',\mu_3,\mu_8,\theta_3,\theta_8)}
{\partial \theta_3^2} \bigg|_{\theta_3 = \theta_8 = 0}, \nonumber \\
\langle W_8^2 \rangle&=&
\frac{\partial^2 \log Z(\beta,L,L',\mu_3,\mu_8,\theta_3,\theta_8)}
{\partial \theta_8^2} \bigg|_{\theta_3 = \theta_8 = 0}.
\end{eqnarray}
Here $Z(\beta,L,L',\mu_3,\mu_8,\theta_3,\theta_8)$ is the partition 
function of a system with twisted periodic boundary conditions in the 
spatial direction of size $L$. The twist is characterized by the matrix
$\exp(i \theta_3 T^3 + i \theta_8 T^8) \in U(1)_3 \times U(1)_8 \subset
SU(3)$. In the limit $L, \beta \rightarrow \infty$ we then obtain the
helicity moduli
\begin{equation}
\Upsilon_3 = \frac{L}{\beta} \langle W_3^2 \rangle, \quad
\Upsilon_8 = \frac{L}{\beta} \langle W_8^2 \rangle.
\end{equation}
A non-vanishing helicity modulus signals strong sensitivity to
the twisted boundary condition and hence the existence of an 
infinite correlation length associated with the Kosterlitz-Thouless 
phenomenon. The precise determination of the helicity moduli
$\Upsilon_3$ and $\Upsilon_8$ requires a careful finite-size
analysis, which will be addressed elsewhere.

\begin{figure}[tbh]
\begin{center}
\includegraphics[width=\columnwidth]{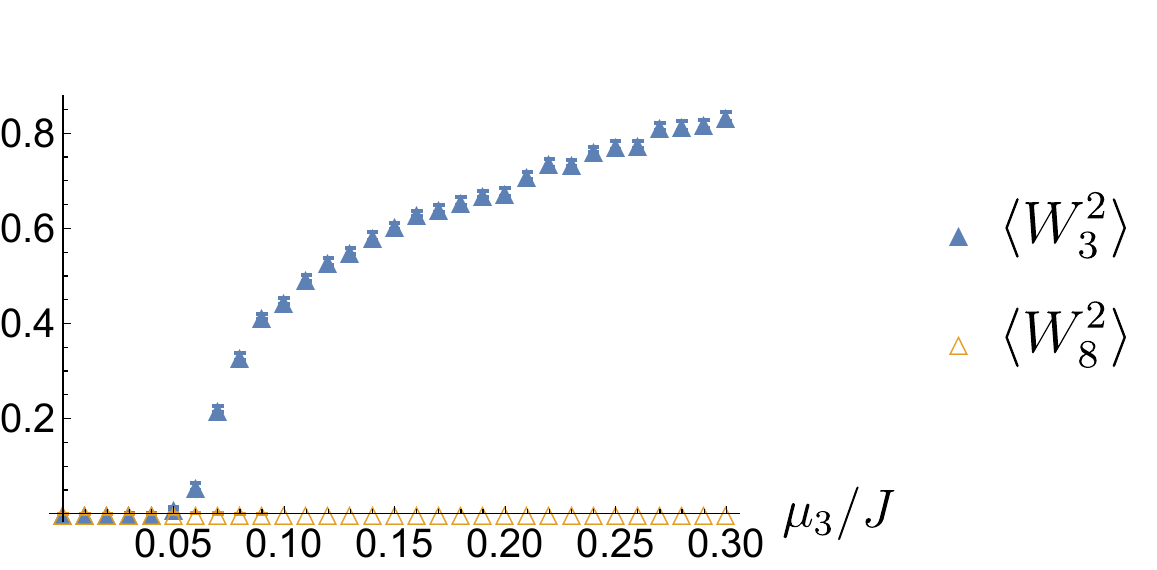}
\caption{\it Expectation values $\langle W_3^2 \rangle$ and
$\langle W_8^2 \rangle$ of the spatial winding numbers $W_3$
and $W_8$ as a function of $\mu_3$ for $\mu_8 = 0$. Lattice
size and inverse temperature are fixed to $L = \beta c = 250 a$.}
\label{W3W8mu3}
\end{center}
\end{figure}

Fig.\ref{W3W8mu3} illustrates the $\mu_3$-dependence of 
$\langle W_3^2 \rangle$ and $\langle W_8^2 \rangle$ for $\mu_8 = 0$ 
with $L = \beta c = 250 a$ (with $\beta J = 140.75$). Since the 
system contains bosons of type $u \overline{d}$ (which have 
$T^8 = 0$), one finds $\langle W_8^2 \rangle = 0$. Since these 
bosons (which appear in the system for $\mu_3 > m c^2$) carry the 
charge $T^3 = 1$, we obtain $\langle W_3^2 \rangle \neq 0$ when 
$\mu_3$ exceeds its critical value. The smooth onset of 
$\langle W_3^2 \rangle$ confirms the second order nature of the 
phase transition. In view of the Mermin-Wagner theorem, we conclude 
that, for $\mu_3 > m c^2$, the bosons of type $u \overline{d}$
undergo the Kosterlitz-Thouless phenomenon. The
``condensation'' affects the $U(1)_3$ symmetry, while the $U(1)_8$
symmetry remains unaffected.

It should be noted that the onset of ``condensation'' is not a
Kosterlitz-Thouless phase transition. While the latter is driven by
thermal fluctuations, below the onset of particle production, the
system just exists in the vacuum state and simply looses its material
basis for ``$\CP(2)$ condensed matter physics''.

\subsection{Double-Species Ferromagnetic 
Bose-Einstein ``Condensation''}

Let us now consider the physics along the $\mu_8$-axis of the phase 
diagram, i.e.\ we now put $\mu_3 = 0$. The system then has an enhanced 
$SU(2)_{ud} \times U(1)_s$ symmetry. As we have seen before, when 
$\mu_8 > \frac{2}{\sqrt{3}} m c^2$, bosons of the two types 
$u \overline{s}$ and $d \overline{s}$ are produced. Again, the question
arises what phase of ``$\CP(2)$ condensed matter'' these bosons form.
Not surprisingly, as we will now show, they form a double-species 
Bose-Einstein ``condensate'', which is now associated with the $U(1)_8$ symmetry.
Interestingly, the $SU(2)_{ud}$ symmetry is realized as
in a ferromagnet. Hence, we address this phase as a double-species
ferromagnetic Bose-Einstein ``condensate''. It should be noted
that in a ferromagnet ``spontaneous symmetry breaking'' is 
qualitatively different than in an antiferromagnet. This is because
the order parameter of the ferromagnet --- namely the uniform 
magnetization or total spin --- is a conserved quantity, while the
staggered magnetization of an antiferromagnet is not conserved. In
particular, this leads to a quadratic dispersion relation of
ferromagnetic spinwaves. This is indeed what happens for the
ferromagnetic Bose-Einstein ``condensate''.

\begin{figure}[tbh]
\begin{center}
\includegraphics[width=0.482\columnwidth]
{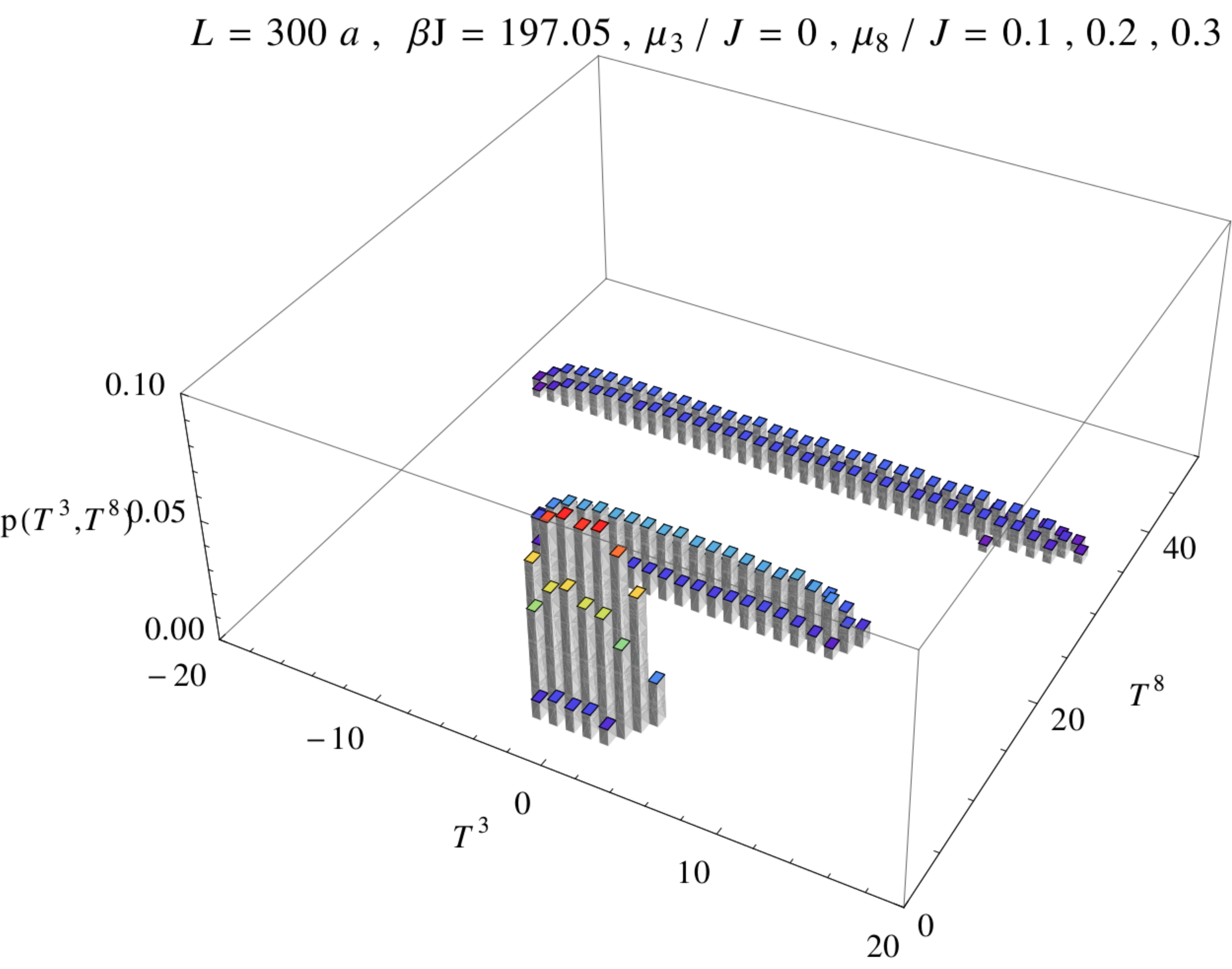} \\
\includegraphics[width=0.482\columnwidth]
{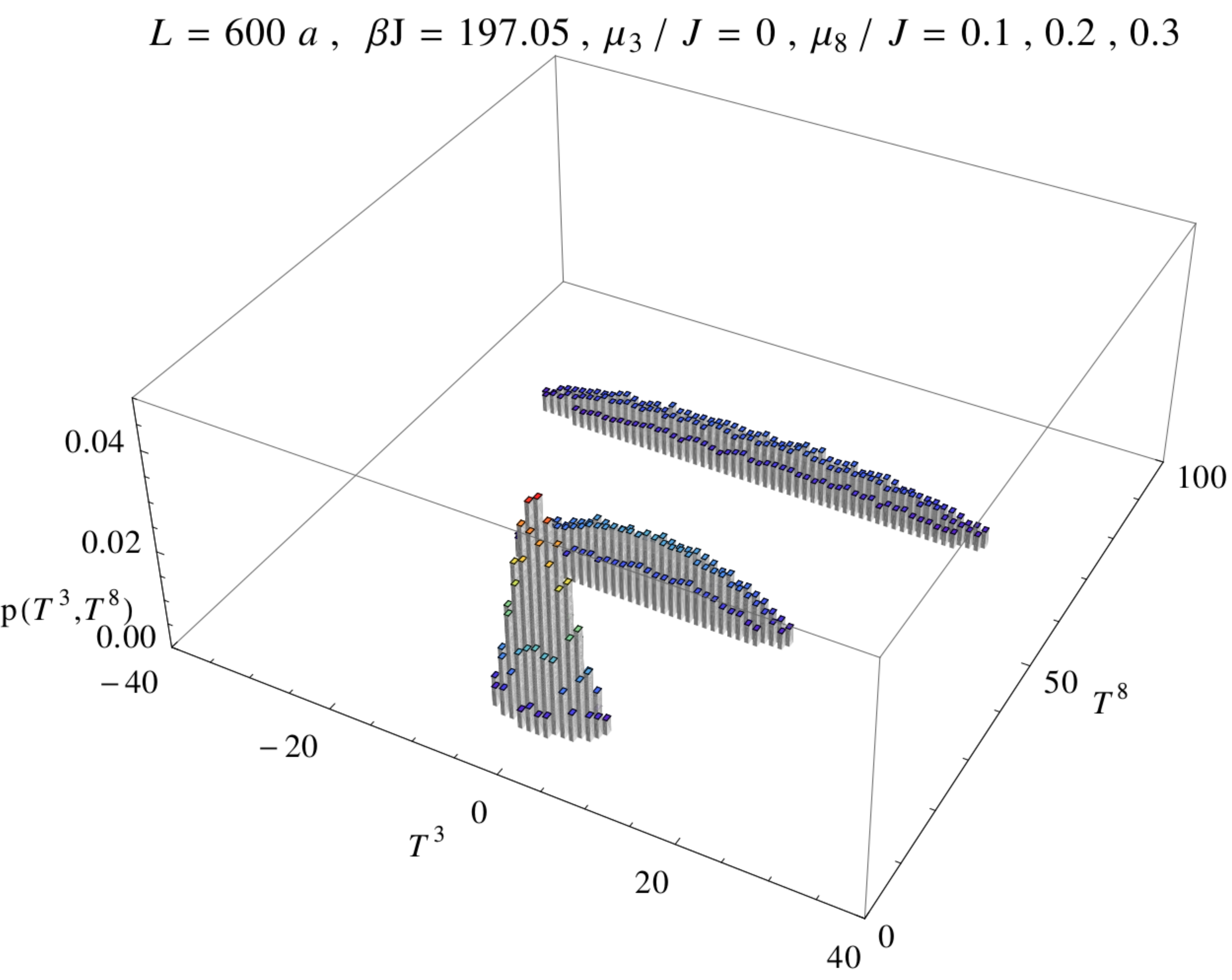} \\
\includegraphics[width=0.482\columnwidth]
{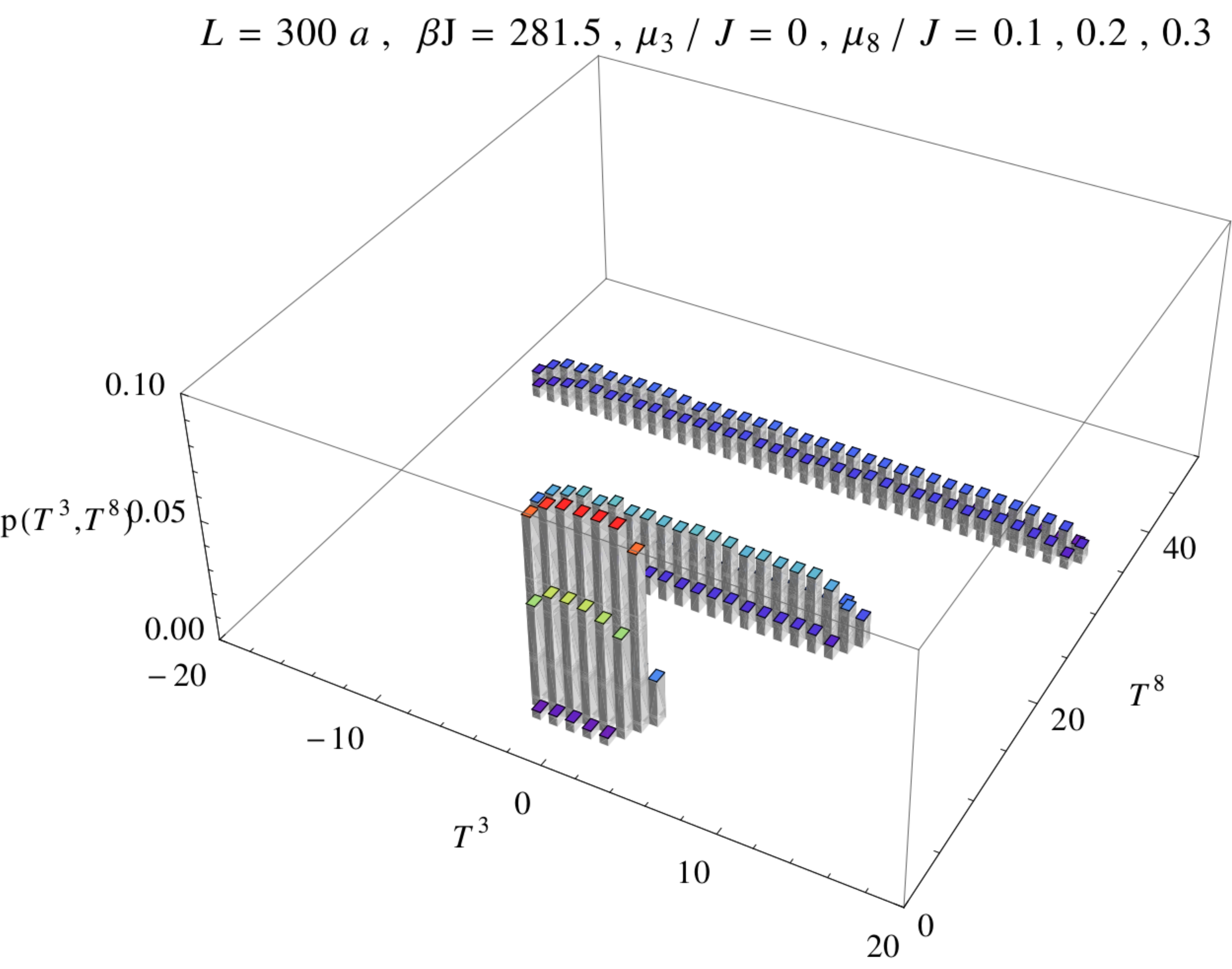}
\caption{\it Probability distributions $p(T^3,T^8)$ for three 
different values of the chemical potential $\mu_8 = 0.1, 0.2, 0.3$
and $\mu_3 = 0$. Lattice size and inverse temperature are fixed to
$L = 300 a$, $\beta J = 197.05$ (top), $L = 600 a$, 
$\beta J = 197.05$ (middle), and $L = 300 a$, $\beta J = 281.5$ 
(bottom).}
\label{histT3T8}
\end{center}
\end{figure}

Fig.\ref{histT3T8} (top) shows the probability distributions of the 
various charge sectors $(T^3,T^8)$ for $\mu_8/J = 0.1, 0.2, 0.3$ 
with $L = \beta c = 300 a$ and $\beta J = 197.05$. Below the threshold 
for particle production (at $\mu_8 = 0.03 J < \frac{2}{\sqrt{3}} m c^2$), 
up to small thermal fluctuations, the system stays in the vacuum sector 
with $T^3 = T^8 = 0$. For $\mu_8 > \frac{2}{\sqrt{3}} m c^2$, on the 
other hand, states with $\langle T^8 \rangle > 0$ are generated.
The most probable values of $\frac{2}{\sqrt{3}} T^8$ for
$\mu_8/J = 0.1, 0.2, 0.3$ are $6, 19, 35$, respectively.
While the thermal fluctuations of $T^8$ are rather small, $T^3$
varies over the whole range $|T^3| \leq T^8/\sqrt{3}$ because bosons 
of both types $u \overline{s}$ and $d \overline{s}$ are equally 
favored by the chemical potential $\mu_8$. Fig.\ref{histT3T8} (middle)
shows data for the larger spatial volume $L = 600 a$, keeping the 
temperature unchanged. As before, for a given value of 
$\mu_8$, about twice as many particles are being produced, but their 
density $\langle T^8\rangle/L$ remains essentially unchanged. This
again indicates that finite-size effects are under control. 
Fig.\ref{histT3T8} (bottom) shows results for the $L = 300 a$ box at 
the lower temperature corresponding to $\beta J = 281.5$. Again, thermal 
fluctuations in the charge distribution are then further suppressed.
We conclude that, as $\mu_8$ increases from $\frac{2}{\sqrt{3}} m c^2$
to $0.3 J$, the system contains an increasing density of bosons of 
type $u \overline{s}$ or $d \overline{s}$.

It is interesting to note that the probability distribution is rather
flat as a function of $T^3$, at least at low temperatures and as long 
as $|T^3| \leq T^8/\sqrt{3}$. This indicates that, in the zero 
temperature limit, the system has a degenerate ground state with a
large value $T = T^8/\sqrt{3}$ for the length of the $SU(2)_{ud}$ 
vector $(T^1,T^2,T^3)$. The total number of bosons of type
$u \overline{s}$ or $d \overline{s}$ (which each have 
$T^8 = \frac{\sqrt{3}}{2}$) is given by $N = \frac{2}{\sqrt{3}} T^8$.
The bosons $u \overline{s}$ and $d \overline{s}$ form a doublet 
with $T = \frac{1}{2}$, and $T^3 = \pm \frac{1}{2}$. Hence
$N$ of these bosons can form a state with maximal total charge
$T = \frac{N}{2} = T^8/\sqrt{3}$, which is indeed what the Monte 
Carlo data indicate. Such a state is totally symmetric under the
permutation of the flavor indices of the $N$ bosons. Hence, their
orbital wave function must also be totally symmetric. This is 
exactly what one expects for a Bose-Einstein condensate. Since the
vector $(T^1,T^2,T^3)$ (just like the total spin of a ferromagnet)
serves as a conserved order parameter for this state, we are 
confronted with a two-component ferromagnetic Bose-Einstein
``condensate''. In this case, the Abelian $U(1)_8 = U(1)_s$ subgroup 
of the $SU(2)_{ud} \times U(1)_s$ symmetry is affected by the 
Kosterlitz-Thouless phenomenon, while the non-Abelian $SU(2)_{ud}$
symmetry is realized ferromagnetically.

\begin{figure}[tbh]
\begin{center}
\includegraphics[width=\columnwidth]{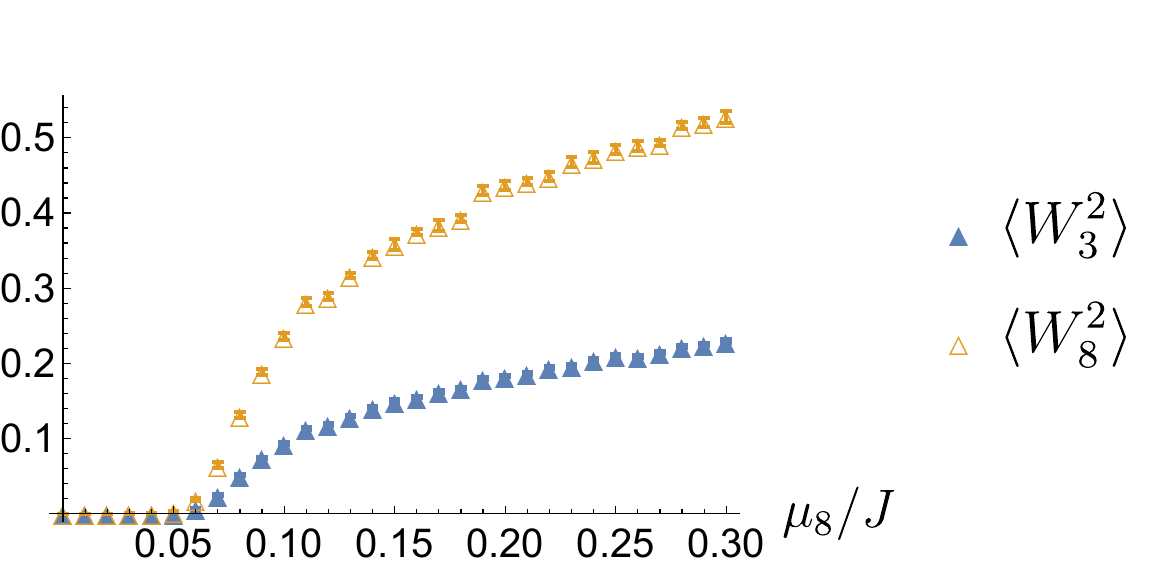}
\caption{\it Expectation values $\langle W_3^2 \rangle$ and
$\langle W_8^2 \rangle$ of the spatial winding numbers $W_3$
and $W_8$ as a function of $\mu_8$ for $\mu_3 = 0$. Lattice
size and inverse temperature are fixed to $L = \beta c = 250 a$.}
\label{W3W8mu8}
\end{center}
\end{figure}

In order to confirm that the bosons indeed ``condense'', we have
again investigated the spatial winding numbers. Fig.\ref{W3W8mu8} 
shows $\langle W_3^2 \rangle$ and $\langle W_8^2 \rangle$ as a 
function of $\mu_8$ for $\mu_3 = 0$. Since the bosons of type 
$u \overline{s}$ and $d \overline{s}$ bosons have 
$T^8 = \frac{\sqrt{3}}{2}$, in this case not only 
$\langle W_3^2 \rangle \neq 0$ but also 
$\langle W_8^2 \rangle \neq 0$ beyond the threshold
$\mu_8 = 2 m c^2/\sqrt{3}$. The smooth onset again indicates the
second order nature of the phase transition. 

\section{Conclusions}

We have investigated the $(1+1)$-d $\CP(2)$ model in an unconventional
regularization, in which the model results via dimensional reduction
from a $(2+1)$-d antiferromagnetic $SU(3)$ quantum spin ladder, which
is particularly well suited for quantum simulation experiments using
ultracold alkaline-earth atoms in an optical superlattice. The 
continuum limit of the dimensionally reduced model is approached by
gradually increasing the transverse extent $L' = n a$ of the ladder.
Here we have considered even values of $n$, which corresponds to the
vacuum angle $\theta = 0$.

Quantum simulation experiments have been proposed to study $\CP(N-1)$
models in real-time as well as at non-zero chemical potential
\cite{Laf15,Laf16} . This
will serve as a test case for the more long-term goal of quantum
simulating QCD in real time or at large baryon chemical potential. 
Due to very severe sign problems, these dynamics are impossible to 
address with Monte Carlo simulations on classical computers. 
Fortunately, the ``condensed matter physics'' of $\CP(N-1)$ models
can be investigated without encountering a sign problem, using an 
efficient worm algorithm applied to the underlying $SU(N)$ quantum
spin system. This can be used to validate future implementations
of $\CP(N-1)$ quantum simulators. This will help to pave the way
towards quantum simulations of $U(N)$ and $SU(N)$ non-Abelian
gauge theories, ultimately including QCD, which can again be 
realized with ultracold alkaline-earth atoms \cite{Ban13} .

Using a worm algorithm for an $SU(3)$ quantum spin ladder, we have 
investigated the phase diagram of the $(1+1)$-d $\CP(2)$ model as a 
function of the two chemical potentials $\mu_3$ and $\mu_8$. The 
vacuum is realized at zero temperature when the chemical potentials 
are below their critical values, which are determined by the mass gap of the 
$SU(3)$ octet of lightest particles. We have concentrated on 
investigating the phase diagram along the $\mu_3$- and $\mu_8$-axes. 
For $\mu_3 > m c^2, \mu_8 = 0$ the bosons of type $u \overline{d}$ 
form a Bose-Einstein ``condensate'', i.e.\ the $U(1)_3$ subgroup of 
the $U(1)_3 \times U(1)_8$ symmetry is affected by the 
Kosterlitz-Thouless phenomenon. Along the $\mu_8$-axis (i.e.\ for
$\mu_3 = 0$), on the other hand, an enhanced symmetry
$SU(2)_{ud} \times U(1)_s \subset SU(3)$ exists. Then for 
$\mu_8 > \frac{2}{\sqrt{3}} m c^2$, bosons of both types 
$u \overline{s}$ and $d \overline{s}$ are equally favored by the
chemical potential. Now the $U(1)_8$ symmetry is affected by
the Kosterlitz-Thouless phenomenon, which gives rise to a
double-species Bose-Einstein ``condensate''. Interestingly, the
$SU(2)_{ud}$ symmetry is realized as in a ferromagnet.
Indeed, the conserved spin vector $(T^1,T^2,T^3)$ picks up a
non-zero expectation value, which means that the system forms a
ferromagnet.

In the future, it would be interesting to further extend the 
investigation of the $\CP(2)$ phase diagram, away from the axes
and to larger values of the chemical potentials. For very large
values of $\mu_3$ and $\mu_8$ the $SU(3)$ quantum spins of the 
underlying antiferromagnetic ladder system will completely
align, forming a trivial saturated state. It will be interesting 
to investigate whether the single- and double-species Bose-Einstein 
``condensates'' persist until saturation, or whether there are
further phases of $\CP(2)$ ``condensed matter'' yet to be 
discovered. Furthermore, it will also be worth studying the phase
diagram for vacuum angle $\theta = \pi$, which corresponds to an
odd number $n$ of transversely coupled quantum spin chains. Then
charge conjugation is spontaneously broken in the vacuum, and it
is interesting to investigate how this affects the phase diagram.

Our study shows that $\CP(N-1)$ models, which share many features 
with QCD, have a rich ``condensed matter'' physics. In contrast to
QCD, fortunately $\CP(N-1)$ models at non-zero chemical potential
can be studied with efficient quantum Monte Carlo simulations on 
classical computers. This can be used to validate future quantum 
simulation experiments of $\CP(N-1)$ models which --- just like 
quantum simulations of $SU(N)$ gauge theories --- can be realized 
with ultracold alkaline-earth atoms in optical lattices. Quantum 
simulation experiments of $\CP(N-1)$ models thus form a natural 
first step towards the ultimate long-term goal of quantum simulating 
QCD.

\section*{Acknowledgments}

We thank W.\ Bietenholz, M.\ Dalmonte, L.\ Fallani, C.\ Laflamme,
H. Mej\'ia-D\'iaz, and Peter 
Zoller for collaboration and interesting discussions on quantum 
simulations of $\CP(N-1)$ models.
The research leading to these results has received funding from the 
Schweizerischer Na\-tio\-nal\-fonds and from the European Research 
Council under the European Union's Seventh Framework Programme 
(FP7/2007-2013)/ ERC grant agreement 339220.

\begin{appendix}

\section{Monte Carlo Method}

To study the $(2+1)$-d antiferromagnetic $SU(3)$ quantum spin ladder
introduced in Section~\ref{sec:su3system} we have implemented a worm
algorithm in discrete Euclidean time. In 
\cite{Cha02} a meron-cluster algorithm was used to solve the 
sign problem at non-zero chemical potential in the $\CP(1) = O(3)$ 
model. Here the $\CP(2)$ model is simulated using a worm algorithm, 
which is capable of updating the system at non-zero chemical potential 
without encountering a sign problem \cite{Eva16}. 
This method is analogous to the $SU(2)$ case studied in \cite{Syl02}.
For simplicity, we discuss the algorithm in discrete Euclidean time
although it is straightforward to implement it directly in the Euclidean 
time continuum
\cite{Bea96}. 

\subsection{Path Integral Representation of the Grand Canonical Partition Function}

In order to construct a discrete Euclidean time path integral for the quantum spin ladder, the Hamiltonian is split into four non-commuting pieces $H = H_1 + H_2 + H_3 + H_4$,
\begin{align}
		H_1 &= J \sum_{x \in A} T^a_x \overline{T}^a_{x+\hat{1}} \,, &
		H_2 &= J \sum_{x \in A} T^a_x \overline{T}^a_{x+\hat{2}} \,, \nonumber \\
		H_3 &= J \sum_{x \in A} T^a_x \overline{T}^a_{x-\hat{1}} \,, &
		H_4 &= J \sum_{x \in A} T^a_x \overline{T}^a_{x-\hat{2}} \,. 
\end{align}
It should be noted that bonds that extend beyond the open boundary must be omitted from the sums in $H_2$ and $H_4$. 
Using the Suzuki-Trotter formula, the partition function takes the form 
\begin{align}
		Z &= \Tr \exp \left(-\beta \left( H - \mu_a T^a \right)	\right)
			= \lim_{\substack{\epsilon \to 0 \\ \epsilon M  = \beta  }}
		\Tr	\left( \prod_{i=1}^4
				\exp \left( -\epsilon  \left( H_i -\frac{\mu_a T^a}{4}  \right) \right)
		\right)^{M}\nonumber \\
		&= \sum_{[f]} \exp \left( -S[f] \right) \, .
\end{align}
Between all transfer matrix factors a complete set of eigenstates $\ket{f_x}$ of $T_x^3$ and $T_x^8$ (with $f_x\in \{u,d,s\}$ on even sites $x\in A$) or $\ket{\bar{f}_y}$ of $\overline{T}_y^3$ and $\overline{T}_y^8$ (with $\bar{f}_y \in \{\bar{u}, \bar{d}, \bar{s}\}$ on odd sites $y \in B$)
is inserted.
This yields a (2+1)-d system whose additional dimension is 
Euclidean time, which extends over $4M$ time-slices.
Since each $H_i$ consists of commuting contributions $h_{xy} =J  T^a_x \overline{T}^a_{y}$,
the Boltzmann weight $\exp(-S[f])$ of a  quantum spin configuration $[f]$ is a product
over space-time plaquettes associated with a nearest-neighbor pair of spins as well as over individual sites at the open boundary
\begin{align}
	\exp(-S[f]) = &\prod_{x \in A, t= 4p-3} 
				W_{f_{x,t}\bar{f}_{x+\hat{1},t}}^{f^{\prime}_{x,t+1} \bar{f}^{\prime}_{x+\hat{1},t + 1}}
				  \prod_{x \in A, t=4p-2} 
				W_{f_{x,t}\bar{f}_{x+\hat{2},t}}^{f^{\prime}_{x,t+1} \bar{f}^{\prime}_{x+\hat{2},t + 1}}
				\nonumber\\
		 	  &\prod_{x \in A, t= 4p-1}  
				W_{f_{x,t}\bar{f}_{x-\hat{1},t}}^{f^{\prime}_{x,t+1} \bar{f}^{\prime}_{x-\hat{1},t + 1}}
                  \prod_{x \in A, t= 4p} 
				W_{f_{x,t}\bar{f}_{x-\hat{2},t}}^{f^{\prime}_{x,t+1} \bar{f}^{\prime}_{x-\hat{2},t + 1}}
				\nonumber \\
				&\prod_{x \in A, x_2 \in \{1,L'/a\}, t=4p}
				W_{f_{x,t}}^{f^{\prime}_{x,t+1} }
				\prod_{y \in B, y_2 \in \{1,L'/a\}, t=4p}
				W_{\bar{f}_{y,t}}^{\bar{f}^{\prime}_{y,t+1} } \, .
\end{align}
The last two products extend over points at the open boundaries in the 2-direction. 
It should again be noted that bonds extending beyond the open boundary must be omitted from the second and fourth product.
The integer $p$ extends from 1 to $M$.

The plaquette weight takes the form
\begin{align}
	W_{f_{x,t}\bar{f}_{x,t}}^{f^{\prime}_{x,t+1} \bar{f}^{\prime}_{y,t + 1}}=
			\matrixel
					{ f_{x,t}  \bar{f}_{y,t} }
			{		\exp \left( -\epsilon \left( h_{xy} -
					\frac{\mu_a T_x^a + \mu_a\overline{T}_y^a }{4}  \right) \right)}
			{f_{x,t+1}^{\prime} \bar{f}_{y,t+1}^{\prime}}\, . 
\end{align}
It is non-negative and has the following non-zero entries for $f \neq f^{\prime}$
\begin{align}
	 W_{f\bar{f}}^{f\bar{f}}
	 	&= 	\exp \left(\frac{\epsilon J}{6} \right)	
		\frac{1}{3} \left(\exp \left(\frac{3  \epsilon J}{2} \right)	+2\right) 
		=: W_A  \,, \nonumber \\
	W_{f\bar{f}}^{f^{\prime}\bar{f}^{\prime}}
		&=\exp \left(\frac{\epsilon J}{6} \right)\frac{1}{3} \left(\exp \left(\frac{3  \epsilon J}{2} \right)	-1\right) 
		=: W_B	\,, \nonumber	\\
W_{f\bar{f}^{\prime}}^{f\bar{f}^{\prime}}
		&=	\exp \left(\frac{\epsilon J}{6} \right)	\exp \left(\frac{\epsilon}{4} \left( 
			\mu_3 t_{f}^3  + \mu_3t_{\bar{f}^\prime}^3  + \mu_8 t_{f}^8 +  \mu_8 t_{\bar{f}^{\prime}}^8 \right) \right)
		=:	W_{f\bar{f}^\prime} \,,
	\end{align}
where $t^3_f,t^8_f, t^3_{\bar{f}},t^8_{\bar{f}}$ correspond to the charges of $f \in\{u,d,s\}$ and $\bar{f} \in\{\bar{u}, \bar{d}, \bar{s}\}$ 
\begin{align}
	t^3_u &=-t^3_{\bar{u}}= \frac{1}{2} \,,  & t^8_u =-t^8_{\bar{u}}&= \frac{1}{2\sqrt{3}} \,,\nonumber	\\
	t^3_d &=-t^3_{\bar{d}}= -\frac{1}{2} \,,& t^8_d =-t^8_{\bar{d}}&= \frac{1}{2\sqrt{3}} \,,	\nonumber\\
	t^3_s &=-t^3_{\bar{s}}= 0			 \,, & t^8_s =-t^8_{\bar{s}}&= -\frac{1}{\sqrt{3}} 	\, .
\end{align}
The weights associated with time-like bonds for points at the open boundary take the form
\begin{align}
  W_{f_{x, t}}^{f_{x, t+1}^\prime} = \matrixel{f_{x,t}}
		  	{\exp \left( \frac{\epsilon\mu_a T_x^a }{4}\right)}
		  		{f_{x,t+1}^{\prime}}\, , \nonumber \\
  W_{\bar{f}_{y, t}}^{\bar{f}_{y, t+1}^\prime} = \matrixel{\bar{f}_{y,t}}
		  	{\exp \left( \frac{\epsilon\mu_a \overline{T}_y^a}{4} \right)}
		  		{\bar{f}_{y,t+1}^{\prime}}\,.
\end{align}
It has non-zero entries only for $f_{x,t} = f'_{x,t+1}$, which are given by
\begin{align}
	W_f^f &= \exp \left( \frac{\epsilon}{4} \left(  \mu_3 t^3_f +  \mu_8 t^8_f\right) \right) =: W_f \,, \\
		W_{\bar{f}}^{\bar{f}} 
		&= \exp \left( \frac{\epsilon}{4} \left(  \mu_3 t^3_{\bar{f}} +  \mu_8 t^8_{\bar{f}}\right) \right) =: W_{\bar{f}} \,.
\end{align}

\subsection{Worm Algorithm}
\label{ssub:worm_algorithm}
After Trotter decomposition, the configurations of the $SU(3)$ quantum spin ladder can be sampled  with a worm algorithm respecting detailed balance and ergodicity.
This algorithm is analogous to the $SU(2)$ case discussed in \cite{Syl02}.
In order to move from one allowed configuration to the next, the worm algorithm proceeds via configurations for which
$SU(3)$ spin conservation is violated at two space-time points associated with the worm-head and the worm-tail.
The worm-head is moved around by a local Metropolis algorithm, until it ultimately meets the tail and the worm closes. 
In that moment spin conservation is again restored and one obtains a new allowed configuration.
By histograming the position of the worm-head relative to the worm-tail one obtains information about the 
two-point-functions of the shift operators $T^{\pm}, U^{\pm}, V^{\pm}$.
In addition, by counting how often the worm-head wraps around the periodic spatial or temporal boundaries
(before it meets the tail) one can determine the changes in the spatial and temporal winding numbers $W_3,W_8$ and $T^3,T^8$.
The worm algorithm proceeds in the following steps:
\begin{enumerate}	
		\item Consider a valid initial configuration $[f^{(0)}]$ of $SU(3)$ quantum spin variables.
		\item Select a space-time point $x_0,t_0$ at random as the initial position of
			   	the worm-head and -tail, as well as an initial time-direction $D=D_0=\pm$.
			Identify the flavor $f=f_{x_0, t_0}$ (or $\bar{f}=\bar{f}_{y_0, t_0}$ 
			for sites $y_0$ on sublattice $B$).  
			Choose a flavor $f^\prime$ different from $f$ at random, and 
			identify the charges carried by the worm as 
			$\Delta^{3,8} = D_0 (t_{f^\prime}^{3,8} - t_{f}^{3,8})$.
	\item Identify the plaquette (or time-like bond for points at the open boundary) at the position $x,t$
			of the worm-head in direction $D$.
		 Choose an exit point $x',t'$ on this plaquette (or time-like bond) according to 
			the probability denoted by $p_{\Delta^{3,8}} \left( x',t' | x,t; \{f\}\right)$ 
			(cf.\ Table \ref{tab:wormrules}), where $\{f\}$ 
			refers to the configuration of the plaquette (or time-like bond) 
			before the move of the worm-head.
			Then move the worm-head to the new position $x',t'$.
			If $x',t'$ agrees with $x,t$ the worm-head bounces, i.e.\ it changes its direction.
			Increment the histogram of the two-point-function of the corresponding shift operators.
			Also record the contribution to the spatial and temporal winding number changes,
		   	according to the direction of the motion of the worm-head.	
	\item Determine the new worm direction $D^\prime$. 
		If $t'=t$, set $D^\prime = - D$, otherwise $D^\prime = D$.
	\item Update the quantum spin at $x',t'$ by adding $D' \Delta^3$ to $t^3_{f_{x',t'}}$ 
			and $D' \Delta^8$ to $t^8_{f_{x',t'}}$.	
			As an example consider, a worm with charges $\Delta^3=1$ and $\Delta^8=0$.
		Moving forward in time, a flavor $d$ will be updated to $u$ and a flavor $\bar{u}$ to $\bar{d}$.
		Moving backward in time, a flavor $u$ will be updated to $d$ and a flavor $\bar{d}$ to $\bar{u}$.
	\item Now replace $D$ by $D'$. If $x',t'$ is different from $x_0,t_0$, i.e.\ as long as the worm-head has not met the tail, proceed with step 3. If $x',t'$ agrees with $x_0,t_0$, i.e.\ if the worm has closed, proceed with step 2 until the desired statistics is achieved. 
\end{enumerate}

The probabilities $p_{\Delta^{3,8}} \left( x',t' | x,t,\{f\} \right)$ are constrained by detailed 
balance and  normalization conditions.
Sylju\aa sen and Sandvik outlined a procedure to state these conditions in the form of several 
decoupled sets of linear equations for a general nearest-neighbor interaction \cite{Syl02}.
Here we extend their $SU(2)$ algorithm to $SU(3)$.
We separately consider each worm-type characterized by the charge that it carries.
The resulting sets of equations for a worm that carries the charges $\Delta^3=-1$ and $\Delta^8=0$ are
shown in Tables \ref{tab:looptab1} and \ref{tab:looptab2} alongside a visualization of
the corresponding worm moves analogous to the ones in \cite{Syl02}.
For worms carrying other charges, the corresponding systems can be obtained by flavor permutations.
In addition to the systems of type $(I)$ (which are of the same form as the ones 
discussed in \cite{Syl02} for $SU(2)$), we have to consider situations where
a worm encounters a plaquette containing the third flavor or an open boundary. 
These lead to the systems of equations of type $(II)$ to $(IV)$ (cf.\ Table \ref{tab:looptab2}).
All systems are under-determined, but can be solved uniquely by imposing the non-negativity 
of all weights and minimizing the sum of all bounce weights $\sum_{i=1}^9( b_i+b_i^{\prime})$, 
which strongly affects the efficiency of the algorithm.

\begin{table}
		\centering
	\input{tables/dirloopeqt1.tex}

	\caption{Worm equations of type $(I)$. Space-time plaquette configurations 
			$_{f_{x,t}\bar{f}_{x,t}}^{f^{\prime}_{x,t+1} \bar{f}^{\prime}_{y,t + 1}}$ 
			are shown together with the path taken by the worm-head. Its entrance and exit 
			points are denoted by a dot and an arrow-head, respectively. The corresponding 
			normalization conditions for the bounce weights $b_i$ and the weights $a, b, c$ for
			space-like, time-like, and diagonal moves are shown on the right.}
	\label{tab:looptab1}
\end{table}

\begin{table}
	\input{tables/dirloopeqt2.tex}

	\caption{Worm equations of type $(II)$, $(III)$, and $(IV)$. Space-time plaquettes 
			(types $(II)$ and $(III)$) and time-like bonds (type $(IV)$) are treated as 
			in Table~\ref{tab:looptab1}.}
	\label{tab:looptab2}
\end{table}

Let us start with a system of equations of type $(I)$ which involves only two flavors 
\begin{align}
	b_1 + a + b &= W_{A} \nonumber\,,	\\
	a + b_2 + c &= W_{B} \nonumber\,,	\\
	b + c + b_3 &= W_{u\bar{d}}	\, .
\end{align}
Here $a$,$b$, and $c$ are the weights for space-like, time-like, and diagonal (space- and time-like)
moves of the worm-head, respectively.
The $b_i$ represent the corresponding bounce weights.
This system can be solved for $a,b,c$ as a function of the bounce weights $b_1,b_2,b_3$ as
\begin{align}
		a &= \frac{1}{2}\left(W_A+W_B-W_{u\bar{d}}\,\,-b_1-b_2+b_3\right) \,,\nonumber \\
		b &= \frac{1}{2}\left(W_A-W_B+W_{u\bar{d}}\,\,-b_1+b_2-b_3\right) \,,\nonumber \\
		c &= \frac{1}{2}\left(-W_A+W_B+W_{u\bar{d}}\,\,+b_1-b_2-b_3\right) \,.
\end{align}

Ideally, we would favor $b_1=b_2=b_3=0$. However, for non-zero chemical potential this
may result in negative weights.
There are three possible cases:
\begin{enumerate}
		\item If $W_A + W_B - W_{u\bar{d}} < 0$,
			the weight $a$ can be made positive by   
			taking $b_3 >0$. This yields the solution
		\begin{align}
			b_1 &= 0  \,,&
			b_2 &= 0  \,, &
			b_3 &= W_{u\bar{d}} - W_A - W_B\,, \nonumber \\
			a &= 0  \,,&
			b &= W_A  \,,&
			c &= W_B \,.
		\end{align} 
			Here all weights are indeed non-negative. This solution minimizes $b_1+b_2+b_3$ 
			for non-negative weights. 
			We cannot choose $b_3$ smaller, since this would yield a negative weight $a$. 
			We have no other way to achieve this, since all $b_i\geq0$. 
		\item 
			If $-W_A + W_B + W_{u\bar{d}} < 0$, the weight $c$ can be made positive by taking $b_1>0$.
		   This implies
			\begin{align}
			b_1 &= W_A - W_B -W_{u\bar{d}} \,, &
			b_2 &= 0 \,,&
			b_3 &= 0 \,, \nonumber\\
			a &= W_B \,, &
			b &= W_{u\bar{d}}\,, &
			c &= 0 \,,
		\end{align} 
			which are again all non-negative.
		   	This solution again minimizes $b_1+b_2+b_3$ according to the same argument as before.
		\item On the other hand, if $W_A + W_B - W_{u\bar{d}} \geq 0$ and 
			   $-W_A + W_B + W_{u\bar{d}} \geq 0$, we can avoid bouncing in this set of equations and obtain
			\begin{align}
			b_1 &=0 \,, &
			b_2 &= 0 \,, &
			b_3 &= 0 \,,  \nonumber\\
			a &= \frac{1}{2}\left(W_A+W_B-W_{u\bar{d}} \right)\,,&
			b &= \frac{1}{2}\left(W_A-W_B+W_{u\bar{d}} \right)\,,\nonumber \\
			c &= \frac{1}{2}\left(-W_A+W_B+W_{u\bar{d}} \right)\,.
			\end{align}

\end{enumerate}
Note that $W_A - W_B + W_{u\bar{d}}$ cannot be negative because $W_A -W_B= \exp(\epsilon J/6)>0$.

The sets of equations of type $	(II)$ correspond to three flavors and are of the form
\begin{align}
		b_4 + d &= W_{u\bar{s}} \nonumber\,, \\
		d + b_5 &= W_{d\bar{s}} \, .
\end{align}
In general $W_{u\bar{s}} \neq W_{d\bar{s}}$ and either $b_4$ or $b_5$ has to be non-zero.
If $W_{u\bar{s}} \geq  W_{d\bar{s}}$,
the solution that minimizes $b_4 + b_5$ is
\begin{align}
		d &=  W_{d\bar{s}}\,,   &
		b_4 &= W_{u\bar{s}}-W_{d\bar{s}} \,, &
		b_5 &= 0\,.
\end{align}
For $W_{u\bar{s}} \leq  W_{d\bar{s}}$, on the other hand, we obtain
\begin{align}
		d &=  W_{u\bar{s}}\,,   &
		b_4 &= 0\,, &
		b_5 &= W_{d\bar{s}}-W_{u\bar{s}} \,.
\end{align}

Sets of equations of type $(III)$ again concern three flavors and are of the form
\begin{align}
		b_6 + e &= W_B \, \,,\nonumber\\
		e + b_7 &= W_B \,.
\end{align}
Here the bounce probabilities can be set to zero and $e = W_B$.

Sets of equations of type ($IV$) are associated with the open boundary
\begin{align}
		b_8 + g &= W_u \, \,,\nonumber\\
		 g + b_9 &= W_d \,.
\end{align}
If $W_{u} \geq  W_{d}$,
the solution that minimizes $b_8 + b_9$ is
\begin{align}
		g &=  W_{d}\,,   &
		b_8 &= W_{u}-W_{d} \,, &
		b_9 &= 0\,.
\end{align}
For $W_{u} \leq  W_{d}$, on the other hand, we obtain
\begin{align}
		g &=  W_{u}\,,   &
		b_8 &= 0\,, &
		b_9 &= W_{d}-W_{u} \,.
\end{align}
Completely analogous solutions exist for the weights $a',b',\ldots,g',b_i'$
as well as for the other worm-types with permuted flavors.

From the above weights for the various possible moves of the worm-head we can now
determine the probabilities $p_{\Delta^{3,8}} \left( x',t'| x,t, \{f\} \right)$,
by normalizing $a,b,\ldots ,g, b_i$ with the corresponding  plaquette or
time-like bond weight (cf.\ Table \ref{tab:wormrules}). 

\begin{table}
		\input{tables/wormrules1p.tex}
		\caption{Rules for a worm carrying the charge $\Delta^3=-1$ and $\Delta^8=0$ 
				in the regimes $(A)$ and $(B)$. Probabilities for moves of the worm-head 
				after entering a space-time plaquette or time-like bond at the entrance point marked by
	   			an arrow. Bounces as well as space-like, time-like, and diagonal moves have the 
				probabilities $p_b, p_{=}, p_{||}, p_{\sf{x}}$, respectively. 
				In this table we have abbreviated $W_{AB} = W_A + W_B$.	}
		\label{tab:wormrules}
\end{table}

\subsection{Algorithmic Phase Diagram}
\label{sub:algorithmic_phase_diagram}
We have seen that the number of non-vanishing bounce weights depends on the values of the chemical potentials.
As a result, for each worm-type we can identify four different chemical potential regimes.
For example, for a worm that carries the charges $\Delta^3=-1$ and $\Delta^8=0$ we distinguish four regimes $(A),(B),(C),(D)$ 
\begin{align}
	   &(A) 
			&& W_{d\bar{u}}\leq 1 ,\, 
			  W_A +W_B \leq W_{u\bar{d}}
			&& \iff 
			&&  \mu_3 \Delta^3 +\mu_8 \Delta^8 \leq -4 J \,,
			\nonumber\\
	   &(B)
			&& W_{d\bar{u}}\leq 1  ,\, 
			 1\leq W_{u\bar{d}} \leq W_A +W_B
			&& \iff 
			&& -4 J \leq  \Delta^3\mu_3 +\Delta^8\mu_8 \leq  0\,,
			\nonumber\\
	   &(C) 
			&& 1\leq W_{d\bar{u}}\leq W_A +W_B  ,\,
			 W_{u\bar{d}} \leq 1 
			&& \iff 
			&& 0 \leq \Delta^3\mu_3 +\Delta^8\mu_8 \leq 4 J\,,
			\nonumber\\
	   &(D) 
			&& W_A+W_B\leq W_{d\bar{u}},\,
		   	W_{u\bar{d}} \leq 1
			&& \iff 
			&& 4 J \leq \Delta^3\mu_3 +\Delta^8\mu_8 \,.
\end{align}

While in regime $(A)$, the worm has three different non-zero bounce probabilities when moving forward in Euclidean time, in regime $(B)$ it has two. 
A worm that carries the charges $\Delta^3 = -1$ and $\Delta^8=0$ does not undergo bounces if $ W_{u\bar{d}} = W_{d\bar{u}}$, i.e.\ if $\mu_3=0$.
All probabilities for regimes $(C)$ and $(D)$ can be obtained by exchanging forward with backward propagation and the weight $ W_{u\bar{d}} $ with $ W_{d\bar{u}}$.
The probabilities for all worm-head moves are continuous across the transitions between different regimes.

\begin{figure}[tbh]
		\centering
			\input{figures/algorithmicphases.tex}
		\caption{Algorithmic phase diagram. 
				The thick black lines ($\mu_3 \Delta^3 + \mu_8 \Delta^8 =0$) 
				mark the transition between regimes $(B)$ and $(C)$ for the worms carrying the 
				charges indicated in the figure, where the bounce probabilities
			   	vanish for the respective worms.
				The thin black lines ($\mu_3 \Delta^3 + \mu_8 \Delta^8 = \pm 4 J $) 
				mark the transitions between regimes $(A)$ and $(B)$ as well as $(C)$ and $(D)$ 
				for the respective worms.
				At $\mu_3 = \mu_8 =0$ all thick lines intersect and, as expected, all bounce probabilities 
			    vanish.
				Inside the hexagon (but not on the lines),
			   	all worms are either in the regime $(B)$ or $(C)$ 
				and thus a total of $3\times2=6$ non-zero bounce probabilities is required. 
				On the thick lines inside the hexagon one worm does not bounce, leaving a total 
				of $4$ non-zero bounce probabilities.
			}
		\label{fig:phases}
\end{figure}
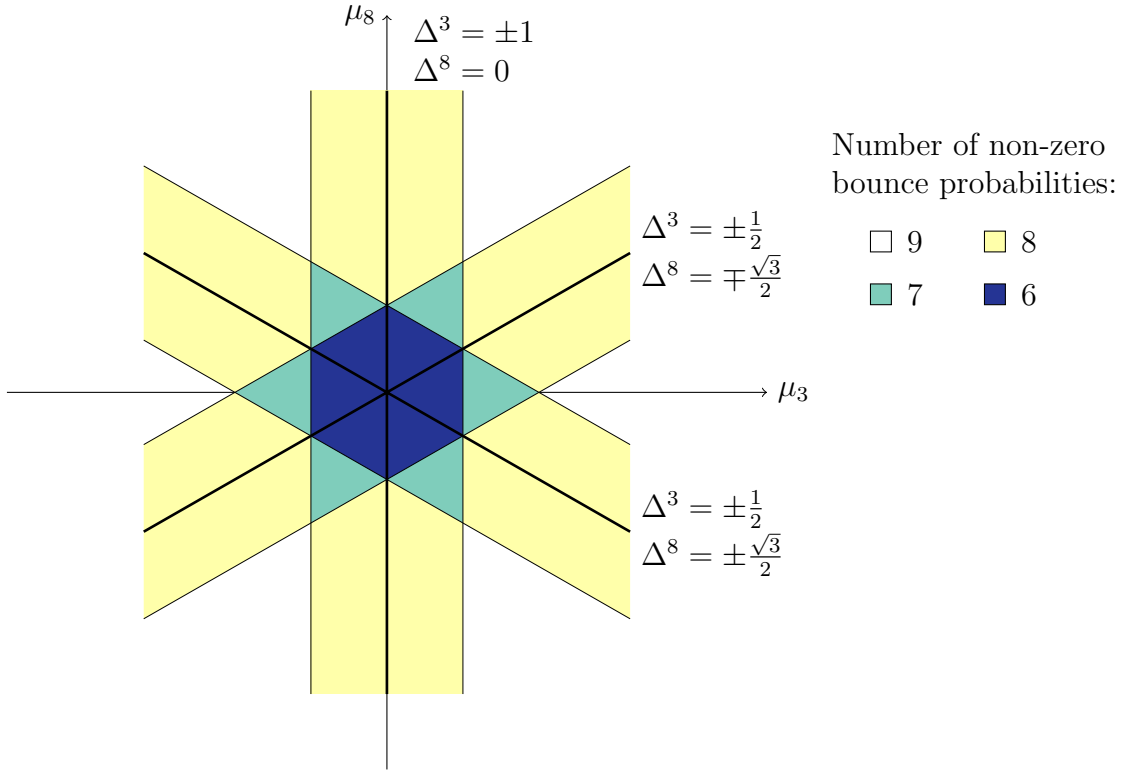

Figure~\ref{fig:phases} shows an ``algorithmic phase diagram'' analogous to Figure 9 in \cite{Syl02}.
Different regimes are distinguished by the number of non-vanishing bounce probabilities.

\subsection{Miscellaneous Coments}
\label{sub:miscellaneous_coments}

Finally we include three miscellaneous comments on the worm algorithm: 
i) Minimizing the probability for diagonal propagation of the worm-head across 
a space-time plaquette instead of minimizing the bounce probabilities does not 
noticeably affect the efficiency of the algorithm. ii) The minimal bouncing 
solution is much more efficient than a heat bath solution of the detailed 
balance relations; in the latter case bouncing probabilities are close to 
50 percent, while in the prior case they are just a few percent. 
iii) Thermalization and autocorrelation times increase with chemical potential, 
but the worm algorithm is still remarkably efficient even for large chemical 
potential.

\end{appendix}

\end{document}

%% file: tables/dirloopeqt1.tex
\newcommand{\stickloop}[3]{
		\begin{tikzpicture}[baseline={([yshift=-.5ex]current bounding box.center)}]
		\def\dx{0.4}
		\def\dy{0.4}
		\def\dex{0.3}
		\def\dey{0.3}
		\def\dtext{0.1}

		\def\ca{(0,0)}
		\def\cc{(0,\dy)}
		\def\cae{(0,0-\dtext)}
		\def\cce{(0,\dy+\dey)}
		\def\cta{(0,0-\dtext)}
		\def\ctc{(0,\dy-\dtext)}

		\node (pe1) at \cae {};
		\node (pe2) at \cce {};

		\node (c1)  at \ca {};
		\node (c2)  at \cc {};

		\draw[thin, gray, fill] (pe#2.center)circle(0.03);
		\draw[->,thick, gray]  (pe#2.center) -- (c#2.center)--(c#3.center)--(pe#3.center);
		\node (p1) [anchor=base] at \cta {$#1$};
		\node (p2) [anchor=base] at \ctc {$#1$};

	\end{tikzpicture}
}

\newcommand{\plaqloop}[6]{
		\begin{tikzpicture}[baseline={([yshift=-.5ex]current bounding box.center)}]
		\def\dx{0.4}
		\def\dy{0.4}
		\def\dex{0.3}
		\def\dey{0.3}
		\def\dtext{0.1}
		
		\def\ca{(0,0)}
		\def\cb{(\dx,0)}
		\def\cc{(0,\dy)}
		\def\cd{(\dx,\dy)}
		
		\def\cta{(0,0-\dtext)}
		\def\ctb{(\dx,0-\dtext)}
		\def\ctc{(0,\dy-\dtext)}
		\def\ctd{(\dx,\dy-\dtext)}

		\def\cae{(0,-\dex)}
		\def\cbe{(\dx,-\dex)}
		\def\cce{(0,\dy+\dey)}
		\def\cde{(\dx,\dy+\dey)}

		\node (pe1) at \cae {};
		\node (pe2) at \cbe {};
		\node (pe3) at \cce {};
		\node (pe4) at \cde {};
		\node (c1)  at \ca {};
		\node (c2)  at \cb {};
		\node (c3)  at \cc {};
		\node (c4)  at \cd {};
		
		\draw[thin, gray, fill] (pe#5.center)circle(0.03);
		\draw[->,thick, gray]  (pe#5.center) -- (c#5.center)--(c#6.center)--(pe#6.center);
		
		\node (p1) [anchor=base] at \cta {$#1$};
		\node (p2) [anchor=base] at \ctb {$\overline{#2}$};
		\node (p3) [anchor=base] at \ctc {$#3$};
		\node (p4) [anchor=base] at \ctd {$\overline{#4}$};
	\end{tikzpicture}
}

\newcommand{\looptablea}[2]{
		\begin{tabular}{ccc}
		\plaqloop{#1}{#1}{#1}{#1}{1}{1}&\plaqloop{#1}{#1}{#1}{#1}{1}{2}&
		\plaqloop{#1}{#1}{#1}{#1}{1}{3}
		\\	
		\plaqloop{#2}{#2}{#1}{#1}{2}{1}&\plaqloop{#2}{#2}{#1}{#1}{2}{2}&
		\plaqloop{#2}{#2}{#1}{#1}{2}{3}
		\\
		\plaqloop{#2}{#1}{#2}{#1}{3}{1}&\plaqloop{#2}{#1}{#2}{#1}{3}{2}&
		\plaqloop{#2}{#1}{#2}{#1}{3}{3}
		\end{tabular}
}

\newcommand{\looptableb}[2]{
		\begin{tabular}{ccc}
		\plaqloop{#1}{#1}{#1}{#1}{2}{2}&\plaqloop{#1}{#1}{#1}{#1}{2}{1}&
		\plaqloop{#1}{#1}{#1}{#1}{2}{4}
		\\	
		\plaqloop{#2}{#2}{#1}{#1}{1}{2}&\plaqloop{#2}{#2}{#1}{#1}{1}{1}&
		\plaqloop{#2}{#2}{#1}{#1}{1}{4}
		\\
		\plaqloop{#1}{#2}{#1}{#2}{4}{2}&\plaqloop{#1}{#2}{#1}{#2}{4}{1}&
		\plaqloop{#1}{#2}{#1}{#2}{4}{4}
		\end{tabular}
}

\newcommand{\looptablec}[2]{
		\begin{tabular}{ccc}
		\plaqloop{#1}{#1}{#1}{#1}{3}{3}&\plaqloop{#1}{#1}{#1}{#1}{3}{4}&
		\plaqloop{#1}{#1}{#1}{#1}{3}{1}
		\\	
		\plaqloop{#1}{#1}{#2}{#2}{4}{3}&\plaqloop{#1}{#1}{#2}{#2}{4}{4}&
		\plaqloop{#1}{#1}{#2}{#2}{4}{1}
		\\
		\plaqloop{#2}{#1}{#2}{#1}{1}{3}&\plaqloop{#2}{#1}{#2}{#1}{1}{4}&
		\plaqloop{#2}{#1}{#2}{#1}{1}{1}
		\end{tabular}
}
\newcommand{\looptabled}[2]{
		\begin{tabular}{ccc}
		\plaqloop{#1}{#1}{#1}{#1}{4}{4}&\plaqloop{#1}{#1}{#1}{#1}{4}{3}&
		\plaqloop{#1}{#1}{#1}{#1}{4}{2}
		\\	
		\plaqloop{#1}{#1}{#2}{#2}{3}{4}&\plaqloop{#1}{#1}{#2}{#2}{3}{3}&
		\plaqloop{#1}{#1}{#2}{#2}{3}{2}
		\\
		\plaqloop{#1}{#2}{#1}{#2}{2}{4}&\plaqloop{#1}{#2}{#1}{#2}{2}{3}&
		\plaqloop{#1}{#2}{#1}{#2}{2}{2}
		\end{tabular}
}

\newcommand{\looptableea}[3]{
\begin{tabular}{cc}
	\plaqloop{#1}{#1}{#2}{#2}{1}{1} &	\plaqloop{#1}{#1}{#2}{#2}{1}{2} \\
	\plaqloop{#3}{#3}{#2}{#2}{2}{2} &	\plaqloop{#3}{#3}{#2}{#2}{2}{1} 
\end{tabular}
}
\newcommand{\looptableeb}[3]{
\begin{tabular}{cc}
	\plaqloop{#1}{#1}{#2}{#2}{2}{2} &	\plaqloop{#1}{#1}{#2}{#2}{2}{1} \\
	\plaqloop{#3}{#3}{#2}{#2}{1}{2} &	\plaqloop{#3}{#3}{#2}{#2}{1}{1} 
\end{tabular}
}
\newcommand{\looptableec}[3]{
\begin{tabular}{cc}
	\plaqloop{#1}{#1}{#2}{#2}{3}{3} &	\plaqloop{#1}{#1}{#2}{#2}{3}{4} \\
	\plaqloop{#1}{#1}{#3}{#3}{4}{3} &	\plaqloop{#1}{#1}{#3}{#3}{4}{4} 
\end{tabular}
}
\newcommand{\looptableed}[3]{
\begin{tabular}{cc}
	\plaqloop{#1}{#1}{#2}{#2}{4}{4} &	\plaqloop{#1}{#1}{#2}{#2}{4}{3} \\
	\plaqloop{#3}{#3}{#2}{#2}{3}{4} &	\plaqloop{#3}{#3}{#2}{#2}{3}{3} 
\end{tabular}
}

\newcommand{\looptablefa}[3]{
\begin{tabular}{cc}
	\plaqloop{#1}{#2}{#1}{#2}{1}{1} &	\plaqloop{#1}{#2}{#1}{#2}{1}{3} \\
	\plaqloop{#3}{#2}{#3}{#2}{3}{1} &	\plaqloop{#3}{#2}{#3}{#2}{3}{3} 
\end{tabular}
}
\newcommand{\looptablefb}[3]{
\begin{tabular}{cc}
	\plaqloop{#1}{#2}{#1}{#2}{2}{2} &	\plaqloop{#1}{#2}{#1}{#2}{2}{4} \\
	\plaqloop{#1}{#3}{#1}{#3}{4}{2} &	\plaqloop{#1}{#3}{#1}{#3}{4}{4} 
\end{tabular}
}
\newcommand{\looptablefc}[3]{
\begin{tabular}{cc}
	\plaqloop{#1}{#2}{#1}{#2}{3}{3} &	\plaqloop{#1}{#2}{#1}{#2}{3}{1} \\
	\plaqloop{#3}{#2}{#3}{#2}{1}{3} &	\plaqloop{#3}{#2}{#3}{#2}{3}{3} 
\end{tabular}
}
\newcommand{\looptablefd}[3]{
\begin{tabular}{cc}
	\plaqloop{#1}{#2}{#1}{#2}{4}{4} &	\plaqloop{#1}{#2}{#1}{#2}{4}{2} \\
	\plaqloop{#1}{#3}{#1}{#3}{2}{4} &	\plaqloop{#1}{#3}{#1}{#3}{2}{2} 
\end{tabular}
}

\newcommand{\looptablega}[2]{
\begin{tabular}{cc}
		\stickloop{#1}{1}{1} & \stickloop{#1}{1}{2}\\
		\stickloop{#2}{2}{1} & \stickloop{#2}{2}{2}\\
\end{tabular}
}

\newcommand{\looptablegb}[2]{
\begin{tabular}{cc}
		\stickloop{#1}{2}{2} & \stickloop{#1}{2}{1}\\
		\stickloop{#2}{1}{2} & \stickloop{#2}{1}{1}\\
\end{tabular}

}

		\begin{tabular}{ccc}
		\toprule
		Type (I)\\
			\looptablea{u}{d} &
			\looptableb{d}{u} &
			$\begin{array}{rl}
					b_1 + a  +b &=W_{A}\\\\
					a + b_2  +c &=W_{B}\\\\
					b + c  +b_3 &=W_{d\overline{u}}
			\end{array}$
			 \\
			 \\
			\looptablec{d}{u} &
			\looptabled{u}{d}&
			$\begin{array}{rl}
					b_1^\prime + a^\prime  +b^\prime &=W_{A}\\\\
					a^\prime + b_2^\prime  +c^\prime &=W_{B}\\\\
					b^\prime + c^\prime  +b_3^\prime &=W_{u\overline{d}}
			\end{array}$
			\\
			\bottomrule
		\end{tabular}

%% file: tables/dirloopeqt2.tex
\newcommand{\stickloop}[3]{
		\begin{tikzpicture}[baseline={([yshift=-.5ex]current bounding box.center)}]
		\def\dx{0.4}
		\def\dy{0.4}
		\def\dex{0.3}
		\def\dey{0.3}
		\def\dtext{0.1}

		\def\ca{(0,0)}
		\def\cc{(0,\dy)}
		\def\cae{(0,0-\dtext)}
		\def\cce{(0,\dy+\dey)}
		\def\cta{(0,0-\dtext)}
		\def\ctc{(0,\dy-\dtext)}

		\node (pe1) at \cae {};
		\node (pe2) at \cce {};

		\node (c1)  at \ca {};
		\node (c2)  at \cc {};

		\draw[thin, gray, fill] (pe#2.center)circle(0.03);
		\draw[->,thick, gray]  (pe#2.center) -- (c#2.center)--(c#3.center)--(pe#3.center);
		\node (p1) [anchor=base] at \cta {$#1$};
		\node (p2) [anchor=base] at \ctc {$#1$};

	\end{tikzpicture}
}

\newcommand{\plaqloop}[6]{
		\begin{tikzpicture}[baseline={([yshift=-.5ex]current bounding box.center)}]
		\def\dx{0.4}
		\def\dy{0.4}
		\def\dex{0.3}
		\def\dey{0.3}
		\def\dtext{0.1}
		
		\def\ca{(0,0)}
		\def\cb{(\dx,0)}
		\def\cc{(0,\dy)}
		\def\cd{(\dx,\dy)}
		
		\def\cta{(0,0-\dtext)}
		\def\ctb{(\dx,0-\dtext)}
		\def\ctc{(0,\dy-\dtext)}
		\def\ctd{(\dx,\dy-\dtext)}

		\def\cae{(0,-\dex)}
		\def\cbe{(\dx,-\dex)}
		\def\cce{(0,\dy+\dey)}
		\def\cde{(\dx,\dy+\dey)}

		\node (pe1) at \cae {};
		\node (pe2) at \cbe {};
		\node (pe3) at \cce {};
		\node (pe4) at \cde {};
		\node (c1)  at \ca {};
		\node (c2)  at \cb {};
		\node (c3)  at \cc {};
		\node (c4)  at \cd {};
		
		\draw[thin, gray, fill] (pe#5.center)circle(0.03);
		\draw[->,thick, gray]  (pe#5.center) -- (c#5.center)--(c#6.center)--(pe#6.center);
		
		\node (p1) [anchor=base] at \cta {$#1$};
		\node (p2) [anchor=base] at \ctb {$\overline{#2}$};
		\node (p3) [anchor=base] at \ctc {$#3$};
		\node (p4) [anchor=base] at \ctd {$\overline{#4}$};
	\end{tikzpicture}
}

\newcommand{\looptablea}[2]{
		\begin{tabular}{ccc}
		\plaqloop{#1}{#1}{#1}{#1}{1}{1}&\plaqloop{#1}{#1}{#1}{#1}{1}{2}&
		\plaqloop{#1}{#1}{#1}{#1}{1}{3}
		\\	
		\plaqloop{#2}{#2}{#1}{#1}{2}{1}&\plaqloop{#2}{#2}{#1}{#1}{2}{2}&
		\plaqloop{#2}{#2}{#1}{#1}{2}{3}
		\\
		\plaqloop{#2}{#1}{#2}{#1}{3}{1}&\plaqloop{#2}{#1}{#2}{#1}{3}{2}&
		\plaqloop{#2}{#1}{#2}{#1}{3}{3}
		\end{tabular}
}

\newcommand{\looptableb}[2]{
		\begin{tabular}{ccc}
		\plaqloop{#1}{#1}{#1}{#1}{2}{2}&\plaqloop{#1}{#1}{#1}{#1}{2}{1}&
		\plaqloop{#1}{#1}{#1}{#1}{2}{4}
		\\	
		\plaqloop{#2}{#2}{#1}{#1}{1}{2}&\plaqloop{#2}{#2}{#1}{#1}{1}{1}&
		\plaqloop{#2}{#2}{#1}{#1}{1}{4}
		\\
		\plaqloop{#1}{#2}{#1}{#2}{4}{2}&\plaqloop{#1}{#2}{#1}{#2}{4}{1}&
		\plaqloop{#1}{#2}{#1}{#2}{4}{4}
		\end{tabular}
}

\newcommand{\looptablec}[2]{
		\begin{tabular}{ccc}
		\plaqloop{#1}{#1}{#1}{#1}{3}{3}&\plaqloop{#1}{#1}{#1}{#1}{3}{4}&
		\plaqloop{#1}{#1}{#1}{#1}{3}{1}
		\\	
		\plaqloop{#1}{#1}{#2}{#2}{4}{3}&\plaqloop{#1}{#1}{#2}{#2}{4}{4}&
		\plaqloop{#1}{#1}{#2}{#2}{4}{1}
		\\
		\plaqloop{#2}{#1}{#2}{#1}{1}{3}&\plaqloop{#2}{#1}{#2}{#1}{1}{4}&
		\plaqloop{#2}{#1}{#2}{#1}{1}{1}
		\end{tabular}
}
\newcommand{\looptabled}[2]{
		\begin{tabular}{ccc}
		\plaqloop{#1}{#1}{#1}{#1}{4}{4}&\plaqloop{#1}{#1}{#1}{#1}{4}{3}&
		\plaqloop{#1}{#1}{#1}{#1}{4}{2}
		\\	
		\plaqloop{#1}{#1}{#2}{#2}{3}{4}&\plaqloop{#1}{#1}{#2}{#2}{3}{3}&
		\plaqloop{#1}{#1}{#2}{#2}{3}{2}
		\\
		\plaqloop{#1}{#2}{#1}{#2}{2}{4}&\plaqloop{#1}{#2}{#1}{#2}{2}{3}&
		\plaqloop{#1}{#2}{#1}{#2}{2}{2}
		\end{tabular}
}

\newcommand{\looptableea}[3]{
\begin{tabular}{cc}
	\plaqloop{#1}{#1}{#2}{#2}{1}{1} &	\plaqloop{#1}{#1}{#2}{#2}{1}{2} \\
	\plaqloop{#3}{#3}{#2}{#2}{2}{2} &	\plaqloop{#3}{#3}{#2}{#2}{2}{1} 
\end{tabular}
}
\newcommand{\looptableeb}[3]{
\begin{tabular}{cc}
	\plaqloop{#1}{#1}{#2}{#2}{2}{2} &	\plaqloop{#1}{#1}{#2}{#2}{2}{1} \\
	\plaqloop{#3}{#3}{#2}{#2}{1}{2} &	\plaqloop{#3}{#3}{#2}{#2}{1}{1} 
\end{tabular}
}
\newcommand{\looptableec}[3]{
\begin{tabular}{cc}
	\plaqloop{#1}{#1}{#2}{#2}{3}{3} &	\plaqloop{#1}{#1}{#2}{#2}{3}{4} \\
	\plaqloop{#1}{#1}{#3}{#3}{4}{3} &	\plaqloop{#1}{#1}{#3}{#3}{4}{4} 
\end{tabular}
}
\newcommand{\looptableed}[3]{
\begin{tabular}{cc}
	\plaqloop{#1}{#1}{#2}{#2}{4}{4} &	\plaqloop{#1}{#1}{#2}{#2}{4}{3} \\
	\plaqloop{#3}{#3}{#2}{#2}{3}{4} &	\plaqloop{#3}{#3}{#2}{#2}{3}{3} 
\end{tabular}
}

\newcommand{\looptablefa}[3]{
\begin{tabular}{cc}
	\plaqloop{#1}{#2}{#1}{#2}{1}{1} &	\plaqloop{#1}{#2}{#1}{#2}{1}{3} \\
	\plaqloop{#3}{#2}{#3}{#2}{3}{1} &	\plaqloop{#3}{#2}{#3}{#2}{3}{3} 
\end{tabular}
}
\newcommand{\looptablefb}[3]{
\begin{tabular}{cc}
	\plaqloop{#1}{#2}{#1}{#2}{2}{2} &	\plaqloop{#1}{#2}{#1}{#2}{2}{4} \\
	\plaqloop{#1}{#3}{#1}{#3}{4}{2} &	\plaqloop{#1}{#3}{#1}{#3}{4}{4} 
\end{tabular}
}
\newcommand{\looptablefc}[3]{
\begin{tabular}{cc}
	\plaqloop{#1}{#2}{#1}{#2}{3}{3} &	\plaqloop{#1}{#2}{#1}{#2}{3}{1} \\
	\plaqloop{#3}{#2}{#3}{#2}{1}{3} &	\plaqloop{#3}{#2}{#3}{#2}{3}{3} 
\end{tabular}
}
\newcommand{\looptablefd}[3]{
\begin{tabular}{cc}
	\plaqloop{#1}{#2}{#1}{#2}{4}{4} &	\plaqloop{#1}{#2}{#1}{#2}{4}{2} \\
	\plaqloop{#1}{#3}{#1}{#3}{2}{4} &	\plaqloop{#1}{#3}{#1}{#3}{2}{2} 
\end{tabular}
}

\newcommand{\looptablega}[2]{
\begin{tabular}{cc}
		\stickloop{#1}{1}{1} & \stickloop{#1}{1}{2}\\
		\stickloop{#2}{2}{1} & \stickloop{#2}{2}{2}\\
\end{tabular}
}

\newcommand{\looptablegb}[2]{
\begin{tabular}{cc}
		\stickloop{#1}{2}{2} & \stickloop{#1}{2}{1}\\
		\stickloop{#2}{1}{2} & \stickloop{#2}{1}{1}\\
\end{tabular}

}

		\begin{tabular}{cccc}
		\toprule	
		Type (II)\\
		\looptablefa{u}{s}{d} &$\begin{array}{rl} b_4 + d&=W_{u\bar{s}}
										\\\\d + b_5 &= W_{d\bar{s}}\end{array}$ &
		\looptablefb{s}{d}{u} &$\begin{array}{rl} b_4^\prime + d^\prime &= W_{s\bar{d}}
										\\\\d^\prime + b_5^\prime &= W_{s\bar{u}}\end{array}$ 
		\\	\midrule
		Type (III)\\
		\looptableea{u}{s}{d} & $\begin{array}{rl} b_6 + e &= W_{B}
												\\\\e + b_7 &= W_{B}\end{array}$ &
		\looptableec{s}{d}{u} & $\begin{array}{rl} b_6^\prime + e^\prime &= W_{B}
												\\\\e^\prime + b_7^\prime &= W_{B}\end{array}$ 
		\\	\midrule
		Type (IV)\\
		\looptablega{u}{d} & $ \begin{array}{rl} b_8 + g &= W_{u}
										\\\\g + b_9 &= W_{d}\end{array} $ &
		\looptablega{\bar{d}}{\bar{u}}&$\begin{array}{rl} b_8^\prime + g^\prime &= W_{\bar{u}}
										\\\\g^\prime + b_9^\prime &= W_{\bar{d}}\end{array} $ \\
		\bottomrule
		\end{tabular}

%% file: tables/wormrules1p.tex
\newcommand{\plaqlooponeleg}[5]{
		\begin{tikzpicture}[baseline={([yshift=-.5ex]current bounding box.center)}]
		\def\dx{0.4}
		\def\dy{0.4}
		\def\dex{0.3}
		\def\dey{0.3}
		\def\dtext{0.1}
		
		\def\ca{(0,0)}
		\def\cb{(\dx,0)}
		\def\cc{(0,\dy)}
		\def\cd{(\dx,\dy)}
		
		\def\cta{(0,0-\dtext)}
		\def\ctb{(\dx,0-\dtext)}
		\def\ctc{(0,\dy-\dtext)}
		\def\ctd{(\dx,\dy-\dtext)}

		\def\cae{(0,-\dex)}
		\def\cbe{(\dx,-\dex)}
		\def\cce{(0,\dy+\dey)}
		\def\cde{(\dx,\dy+\dey)}

		\node (pe1) at \cae {};
		\node (pe2) at \cbe {};
		\node (pe3) at \cce {};
		\node (pe4) at \cde {};
		\node (c1)  at \ca {};
		\node (c2)  at \cb {};
		\node (c3)  at \cc {};
		\node (c4)  at \cd {};
		
		\draw[thin, gray, fill] (pe#5.center)circle(0.03);
		\draw[->,thick, gray]  (pe#5.center) -- (c#5.center);
		
		\node (p1) [anchor=base] at \cta {$#1$};
		\node (p2) [anchor=base] at \ctb {$\bar{#2}$};
		\node (p3) [anchor=base] at \ctc {$#3$};
		\node (p4) [anchor=base] at \ctd {$\bar{#4}$};
	\end{tikzpicture}
}

\newcommand{\sticklooponeleg}[3]{
		\begin{tikzpicture}[baseline={([yshift=-.5ex]current bounding box.center)}]
		\def\dx{0.4}
		\def\dy{0.4}
		\def\dex{0.3}
		\def\dey{0.3}
		\def\dtext{0.1}
		
		\def\ca{(0,0)}
		\def\cc{(0,\dy)}
		
		\def\cta{(0,0-\dtext)}
		\def\ctc{(0,\dy-\dtext)}

		\def\cae{(0,-\dex)}
		\def\cce{(0,\dy+\dey)}

		\node (pe1) at \cae {};
		\node (pe3) at \cce {};
		\node (c1)  at \ca {};
		\node (c3)  at \cc {};
		
		\draw[thin, gray, fill] (pe#3.center)circle(0.03);
		\draw[->,thick, gray]  (pe#3.center) -- (c#3.center);
		
		\node (p1) [anchor=base] at \cta {$#1$};
		\node (p3) [anchor=base] at \ctc {$#2$};
	\end{tikzpicture}
}

\begin{tabular}{cccccccccc}
		\toprule
	&&	\multicolumn{4}{l}{Regime $(A)$}
	&	\multicolumn{4}{l}{Regime $(B)$}\\
	&	& $p_b$ & $p_{=}$ & $p_{||}$ & $p_{\sf{X}}$ 
		& $p_b$ & $p_{=}$ & $p_{||}$ & $p_{\sf{X}}$ \\
		\cmidrule(lr){3-6}
		\cmidrule(lr){7-10}
		\plaqlooponeleg{x}{u}{u}{u}{1}&	\plaqlooponeleg{d}{d}{d}{d}{2}
		 	& \(  \frac{1-W_{d\bar{u}}}{W_{A}} \) 
		  	& \(  \frac{W_{B}}{W_{A}}\)
		  	& \(  \frac{W_{d\bar{u}} }{ W_{A}} \)
			& \(  0 \)
		 	& \(  \frac{1-W_{d\bar{u}}}{W_{A}} \) 
		  	& \(  \frac{W_{B}}{W_{A}}\)
		  	& \(  \frac{W_{d\bar{u}} }{ W_{A}} \)
			& \(  0 \)
		\\ 
		\plaqlooponeleg{u}{u}{d}{d}{1}&	\plaqlooponeleg{d}{d}{u}{u}{2}
			& \(   0 \) 
		  	& \(  1 \) 
			& \(  0 \)
	   	  	& \(   0\)
			& \(  0 \) 
		  	& \(  1 \) 
			& \(  0 \)
	   	  	& \(   0\)
		\\	
		\plaqlooponeleg{u}{d}{u}{d}{1}&	\plaqlooponeleg{u}{d}{u}{d}{2}
			& \(   \frac{W_{u\bar{d}} -W_{AB}}{W_{u\bar{d}}} \) 
			& \(  0 \)
		  	& \(  \frac{W_{A}}{W_{u\bar{d}}} \) 
	   	  	& \(   \frac{W_{B}}{W_{u\bar{d}}}\)
			& \(  0 \) 
			& \(  0 \)
		  	& \(   \frac{1 + W_{u\bar{d}}}{2W_{u\bar{d}}}\) 
	   	  	& \(   \frac{-1 +W_{u\bar{d}}}{2W_{u\bar{d}}}\)
		\\
		\plaqlooponeleg{u}{s}{u}{s}{1}&	\plaqlooponeleg{s}{d}{s}{d}{2}
			& \(   1 - \frac{W_d}{W_u} \) 
			& \(  0 \)
	   	  	& \(   \frac{W_d}{W_u}\)
			& \(  0 \)
			& \(  1 - \frac{W_d}{W_u} \) 
			& \(  0 \)
	   	  	& \(   \frac{W_d}{W_u}\)
			& \(  0 \)
		\\
		\sticklooponeleg{u}{u}{1}&	\sticklooponeleg{\bar{d}}{\bar{d}}{1}
			& \(   1 - \frac{W_d}{W_u} \) 
			& \(  0 \)
	   	  	& \(   \frac{W_d}{W_u}\)
			& \(  0 \)
			& \(  1 - \frac{W_d}{W_u} \) 
			& \(  0 \)
	   	  	& \(   \frac{W_d}{W_u}\)
			& \(  0 \)

		\\
		\plaqlooponeleg{u}{u}{s}{s}{1}&	\plaqlooponeleg{d}{d}{s}{s}{2}
			& \(   0 \) 
		  	& \(  1 \) 
			& \(  0 \)
	   	  	& \(   0\)
			& \(  0 \) 
		  	& \(  1 \) 
			& \(  0 \)
	   	  	& \(   0\)
		\\
		\plaqlooponeleg{u}{u}{u}{u}{4}&	\plaqlooponeleg{d}{d}{d}{d}{3}
		  	& \(   0 \) 
		  	& \(  0 \) 
	   	  	& \(   1\)
			& \(  0 \)
		  	& \(  0 \) 
		  	& \(   \frac{W_{AB} - W_{u\bar{d}}}{2W_{A}} \) 
	   	  	& \(   \frac{1 + W_{u\bar{d}}}{2W_{A}} \)
			& \(  0 \)
		\\
		\plaqlooponeleg{u}{u}{d}{d}{3}&	\plaqlooponeleg{d}{d}{u}{u}{4}
			& \(   0 \) 
		  	& \(  0 \) 
			& \(  0 \)
	   	  	& \(   1\)
			& \(  0 \) 
		  	& \(   \frac{W_{AB} - W_{u\bar{d}}}{2W_=} \) 
			& \(  0 \)
	   	  	& \(   \frac{-1 + W_{u\bar{d}}}{2W_=}\)
		\\
		\plaqlooponeleg{d}{u}{d}{u}{3}&	\plaqlooponeleg{d}{u}{d}{u}{4}
			& \(   0 \) 
			& \(  0 \)
		  	& \(  1 \) 
	   	  	& \(   0\)
			& \(  0 \) 
			& \(  0 \)
		  	& \(  1 \) 
	   	  	& \(   0\)
		\\
		\plaqlooponeleg{d}{s}{d}{s}{3}&	\plaqlooponeleg{s}{u}{s}{u}{4}
			& \(   0 \) 
			& \(  0 \)
	   	  	& \(   1\)
			& \(  0 \)
			& \(  0 \) 
			& \(  0 \)
	   	  	& \(   1\)
			& \(  0 \)
	\\
		\sticklooponeleg{d}{d}{3}	&\sticklooponeleg{\bar{u}}{\bar{u}}{3}
			& \(   0 \) 
			& \(  0 \)
	   	  	& \(   1\)
			& \(  0 \)
			& \(  0 \) 
			& \(  0 \)
	   	  	& \(   1\)
			& \(  0 \)
		\\
		\plaqlooponeleg{s}{s}{u}{u}{3} &	\plaqlooponeleg{s}{s}{d}{d}{4}
			& \(   0 \) 
		  	& \(  1 \) 
			& \(  0 \)
	   	  	& \(   0\)
			& \(  0 \) 
		  	& \(  1 \) 
			& \(  0 \)
	   	  	& \(   0\)
		\\
		\bottomrule

\end{tabular}

%% file: figures/algorithmicphases.tex
\begin{tikzpicture}
\draw [->] (-5,0)--(0,0)--(5,0) node[right] {$\mu_3$};
\draw [->] (0,-5)--(0,0)--(0,5) node[left] {$\mu_8$};

\coordinate (h1) at (0,{2/sqrt(3)});
\coordinate (h2) at (1,{1/sqrt(3)});
\coordinate (h3) at (1,{-1/sqrt(3)});
\coordinate (h4) at (0,{-2/sqrt(3)});
\coordinate (h5) at (-1,{-1/sqrt(3)});
\coordinate (h6) at (-1,{1/sqrt(3)});

\coordinate (s1) at (1,{sqrt(3)});
\coordinate (s2) at (2,0);
\coordinate (s3) at (1,{-sqrt(3)});
\coordinate (s4) at (-1,{-sqrt(3)});
\coordinate (s5) at (-2,0);
\coordinate (s6) at (-1,{sqrt(3)});
\coordinate (l1b) at (0,4);
\coordinate (l1e) at (0,-4);
\coordinate (l2b) at (1,4);
\coordinate (l2e) at (1,-4);
\coordinate (l3b) at (-1,4);
\coordinate (l3e) at (-1,-4);

\def\sizex{3.2}

\coordinate (l4b) at (\sizex,{\sizex/sqrt(3) + 0});
\coordinate (l5b) at (\sizex,{\sizex/sqrt(3) + 2/sqrt(3)});
\coordinate (l6b) at (\sizex,{\sizex/sqrt(3) - 2/sqrt(3)});
\coordinate (l4e) at (-\sizex,{-\sizex/sqrt(3) + 0});
\coordinate (l5e) at (-\sizex,{-\sizex/sqrt(3) + 2/sqrt(3)});
\coordinate (l6e) at (-\sizex,{-\sizex/sqrt(3) - 2/sqrt(3)});

\coordinate (l7b) at (\sizex,{-\sizex/sqrt(3) + 0});
\coordinate (l8b) at (\sizex,{-\sizex/sqrt(3) + 2/sqrt(3)});
\coordinate (l9b) at (\sizex,{-\sizex/sqrt(3) - 2/sqrt(3)});
\coordinate (l7e) at (-\sizex,{\sizex/sqrt(3) + 0});
\coordinate (l8e) at (-\sizex,{\sizex/sqrt(3) + 2/sqrt(3)});
\coordinate (l9e) at (-\sizex,{\sizex/sqrt(3) - 2/sqrt(3)});

\definecolor{c0}{rgb}{0.0, 0.0, 0.0}
\definecolor{c4}{rgb}{0.0, 0.0, 0.0}
\definecolor{c6}{RGB}{37,52,148}
\definecolor{c7}{RGB}{127,205,187}
\definecolor{c8}{RGB}{255,255,170}
\definecolor{c9}{RGB}{255,255,255}


\tikzset{b4s/.append style={line width = 1, color=c4}}
\tikzset{b6s/.append style={line width = 1, color=c4}}
\tikzset{transs/.append style={line width = 0.2, color=c4}}

\fill [c8] (l2b)--(l3b)--(l3e)--(l2e)--cycle;
\fill [c8] (l5b)--(l6b)--(l6e)--(l5e)--cycle;
\fill [c8] (l8b)--(l9b)--(l9e)--(l8e)--cycle;
\fill [c7] (h1)--(s1)--(h2)--(s2)--(h3)--(s3)--(h4)--(s4)--(h5)--(s5)--(h6)--(s6)--cycle;
\fill [c6] (h1)--(h2)--(h3)--(h4)--(h5)--(h6)--cycle;

\draw [b4s] (h4)--(0,0) --(h1);
\draw [b4s] (h5)--(0,0) --(h2);
\draw [b4s] (h6)--(0,0) --(h3);

\draw [transs] (l2b)--(l2e); 
\draw [transs] (l3b)--(l3e); 
\draw [transs] (l5b)--(l5e); 
\draw [transs] (l6b)--(l6e); 
\draw [transs] (l8b)--(l8e); 
\draw [transs] (l9b)--(l9e); 

\draw [b6s] (h1)--(l1b);
\draw [b6s] (h2)--(l4b);
\draw [b6s] (h3)--(l7b);
\draw [b6s] (h4)--(l1e);
\draw [b6s] (h5)--(l4e);
\draw [b6s] (h6)--(l7e);
\draw [fill] (0,0) circle (0.03);

\node [right] at ($(l1b) +(0.2,0.8) $) {$\Delta^3 = \pm 1$};
\node [right] at ($(l1b) +(0.2,0.3) $) {$\Delta^8 = 0$};

\node [right] at ($(l4b) +(0,0.3) $) {$\Delta^3 = \pm \frac{1}{2}$};
\node [right] at ($(l4b) +(0,-0.3) $) {$\Delta^8 = \mp \frac{\sqrt{3}}{2} $};

\node [right] at ($(l7b) +(0,0.3) $) {$\Delta^3 = \pm \frac{1}{2}$};
\node [right] at ($(l7b) +(0,-0.3) $) {$\Delta^8 = \pm \frac{\sqrt{3}}{2} $};




\def\dlg{0.7}
\def\dlgx{0.2}

\coordinate (lg1) at (6.5,  2);
\coordinate (lg2) at (8, 2);
\coordinate (lg3) at (6.5, {2-\dlg});
\coordinate (lg4) at (8, {2-\dlg});

\node [right] at ($(lg1)+(-1,1)$) {\begin{tabular}{l} Number of non-zero\\ bounce probabilities: \end{tabular}};
\node [draw,  fill, fill=c9] at (lg1){};
\node [right] at ($(lg1)+(\dlgx,0) $) {9 };
\node [draw,  fill, fill=c8] at (lg2){};
\node [right] at ($(lg2)+(\dlgx,0) $) {8 };
\node [draw,  fill, fill=c7] at (lg3){};
\node [right] at ($(lg3)+(\dlgx,0) $) {7};
\node [draw,  fill, fill=c6] at (lg4){};
\node [right] at ($(lg4)+(\dlgx,0) $) {6 };

\end{tikzpicture}